\documentclass[conference]{IEEEtran}
\IEEEoverridecommandlockouts
\usepackage{xcolor}
\usepackage{soul}
\usepackage{algpseudocode} % algorithm
\usepackage{algorithm}% algorithm
\usepackage{graphicx} % Required for inserting images
\usepackage{comment}
\usepackage{subfiles}
\usepackage{hyperref}
\usepackage{subcaption}
\usepackage{booktabs}
\usepackage{multirow}
\usepackage[most]{tcolorbox}
\usepackage{enumitem}
\usepackage{stfloats}
\usepackage{ragged2e}
\usepackage[utf8]{inputenc}

\def\BibTeX{{\rm B\kern-.05em{\sc i\kern-.025em b}\kern-.08em
    T\kern-.1667em\lower.7ex\hbox{E}\kern-.125emX}}
\usepackage{booktabs} % For formal tables

\usepackage{biblatex} %Imports biblatex package
\begin{document}
\NewDocumentCommand{\mourad}{ mO{} }{\textcolor{blue}{\textsuperscript{\textit{Mourad}}\textsf{\textbf{\small[#1]}}}}
\newcommand{\fb}[1]{\textcolor{red}{FB: #1}}  %farah's comments
\newcommand{\reviewer}[1]{\vspace{2mm}\noindent\textit{{{#1 }}} }

\title{OMNIA: Closing the Loop by Leveraging LLMs for Knowledge Graph Completion}

\author{
\IEEEauthorblockN{
Frédéric Ieng\IEEEauthorrefmark{1},
Soror Sahri\IEEEauthorrefmark{1},
Mourad Ouzzani\IEEEauthorrefmark{2},
Massinissa Hammaz\IEEEauthorrefmark{1}\\
Salima Benbernou\IEEEauthorrefmark{1},
Hanieh Khorashadizadeh\IEEEauthorrefmark{3},
Sven Groppe\IEEEauthorrefmark{3}\IEEEauthorrefmark{4},
Farah Benamara\IEEEauthorrefmark{5}
}

\IEEEauthorblockA{\IEEEauthorrefmark{1}LIPADE, Université Paris Cité, Paris, France}
\IEEEauthorblockA{\IEEEauthorrefmark{2}Qatar Computing Research Institute, HBKU, Doha, Qatar}
\IEEEauthorblockA{\IEEEauthorrefmark{3}Universität zu Lübeck, Germany}
\IEEEauthorblockA{\IEEEauthorrefmark{4}TU Bergakademie Freiberg, Germany}
\IEEEauthorblockA{\IEEEauthorrefmark{5}Université de Toulouse, IRIT, France\\
and IPAL-CNRS, Singapore}

\IEEEauthorblockA{
Emails: \{frederic.ieng, soror.sahri, salima.benbernou\}@u-paris.fr,
massinissa.hammaz@etu.u-paris.fr, mouzzani@hbku.edu.qa,\\
\{hanieh.khorashadizadeh, sven.groppe\}@uni-luebeck.de,
farah.benamara@irit.fr
}
}

\maketitle

\begin{abstract}
Knowledge Graphs (KGs) are widely used to represent structured knowledge, yet their automatic construction, especially with Large Language Models (LLMs), often results in incomplete or noisy outputs. Knowledge Graph Completion (KGC) aims to infer and add missing triples, but most existing methods either rely on structural embeddings that overlook semantics or  language models that ignore the graph’s structure and depend on external sources. In this work, we present OMNIA, a two-stage approach that bridges structural and semantic reasoning for KGC. It first generates candidate triples by clustering semantically related entities and relations within the KG, then validates them through lightweight embedding filtering followed by LLM-based semantic validation. OMNIA performs on the internal KG, without external sources, and specifically targets implicit semantics that are most frequent in LLM-generated graphs.  Extensive experiments on multiple datasets 
demonstrate that OMNIA significantly improves F1-score
%including FB15k-237, CoDEx-M, WN18RR, and a socio-economic LLM-generated KG, show that OMNIA achieves up
%up to 23 percentage points higher F1-score 
compared to traditional embedding-based models.
%reaching an F1-score of 0.91 on CoDEx-M and 0.86 on FB15k-237. 
These results highlight OMNIA’s effectiveness and efficiency, as its clustering and filtering stages reduce both search space and validation cost while maintaining high-quality completion.

\end{abstract}

%
% % The code below should be generated by the tool at
% % http://dl.acm.org/ccs.cfm
% % Please copy and paste the code instead of the example below. 

% \keywords{ACM proceedings, \LaTeX, text tagging}


\section{Introduction}
\label{sec:intro}

\noindent \textbf{\textit{Background.}}
%A Knowledge Graph is a structured representation of human knowledge, connecting different entities through relationships. 
%These entities represent either real-world objects or abstract concepts. 
KGs are structured representations of human knowledge, linking entities—real-world objects or abstract concepts—through relations. Their versatility makes them essential in applications such as search, question answering, and intelligent assistants. By structuring knowledge explicitly, KGs enable machines to reason, infer and retrieve information more effectively. However, as they expand in size and complexity, maintaining high quality becomes crucial to ensure reliable performance.

To assess the quality of a KG, various quality dimensions have been proposed~\cite{zaveri2016quality,WANG2021607}, with particular emphasis on 
(i)~accuracy--correctness of facts and relationships, as inaccuracies can undermine both trust and utility; 
(ii)~completeness--the extent to which all relevant information is captured, since missing data may lead to overlooked insights; and 
(iii)~consistency--logical coherence by avoiding contradictory statements or violations of predefined rules. 
In this paper, we focus on inferring triples that are missing from a given KG, a task that pertains to the completeness dimension. %\fb{why? this is not motivated}

%Several studies have addressed the completeness dimension by proposing methods to assess or improve KG's completeness

Several methods have been proposed to assess or improve KG's completeness  
using external  text corpora~\cite{10912943}, 
rule mining techniques~\cite{meilicke2020reinforcedanytimerulelearning}, or 
embedding-based predictions~\cite{DBLP:conf/nips/BordesUGWY13, trouillon2016complexembeddingssimplelink, sun2019rotateknowledgegraphembedding, yang2015embeddingentitiesrelationslearning, geng2023relational}. These approaches are generally based on fixed ontologies with predefined entities and relations, which limits their ability to handle triples expressed in natural language and implicit semantics, and often ignore the structural relationships encoded within the KG itself.

Large Language Models (LLMs) have recently been applied to KG construction \cite{zhang-etal-2024-fine-tuning, zhu2024llmsknowledgegraphconstruction, trajanoska2023enhancingknowledgegraphconstruction, ye2025constructionapplicationmaterialsknowledge} and completion \cite{li2024contextualization, xu2024multi, shu2024knowledge, 10912943, guo2025ontology}, offering new opportunities to address the limitations of traditional KG completion approaches. However, LLMs still struggle to extract or infer triples that capture implicit semantics and often lack sufficient structural context, which reduces the effectiveness of conventional completion methods on LLM-generated KGs.

\noindent \textbf{\textit{Motivation. }} We consider a real-world use case from a collaborative project aiming to analyze the global response to the COVID-19 pandemic using enriched KGs~\cite{qualityont}. The following example highlights completion challenges that underscore the need for a more effective, semantic-based approach for KGs. Beyond these challenges, our approach specifically targets semantic issues that are prevalent in LLM-generated KGs, such as handling implicit semantics.
%The following example highlights the limitations of existing KG completion methods and motivates the need of a more effective and semantically aware approach, designed to be applicable to any KG. 
%\mourad{Do we really need the following sentence?}
%Beyond addressing general completion challenges, our approach specifically targets issues unique to LLM-generated KGs, such as implicit semantics.}

%We illustrate these challenges and the potential of LLM-enhanced completion through a motivating example drawn from a collaborative project aiming to analyze the global response to the COVID-19 pandemic through enriched, multilingual knowledge graphs.

% The overarching project focuses on capturing how facts evolve across time and languages, identifying inconsistencies and complementarities across diverse data sources such as scientific publications, news, and social media. Within this context, our Goal is to enhance the extraction of implied and reasoned knowledge that LLMs often miss, thereby improving the overall quality and completeness of the resulting knowledge graphs. This task involves generating new, previously unstated triples to enrich the knowledge graph—an area that has received relatively limited attention in existing research.
%\paragraph{\textbf{Example.}} 
\noindent \textit{\textbf{Running example. }}We illustrate this with a COVID-19 KG~\cite{ezzabady2024towards}, generated through an LLM-based extraction process applied to COVID-Fact, a dataset of manually verified facts about the pandemic \cite{saakyan2021covid}. The resulting domain-specific KG captures key information covering regulations, policies, statistics and drug-related effects.
From the following factual statements:
%$f_1$, $f_2$ and $f_3$: 
\begin{itemize}
    \item $f_1$: "Remdesivir and chloroquine effectively inhibit the recently emerged 2019-ncov in vitro"
    \item $f_2$: "FDA approves the emergency use of chloroquine phosphate and hydroxychloroquine sulfate for treatment of COVID-19"
    \item $f_3$: "A chicago hospital treating severe covid-19 patients with gilead sciences ' antiviral medication remdesivir is seeing recoveries in patients ' symptoms , stat news reports"
\end{itemize}
using OpenAI GPT4 ~\cite{ezzabady2024towards}, the following triples were extracted: $t_1$: (remdesivir, treats, sars-cov-2), $ t_2$:
(remdesivir, inhibits, 2019-ncov), and $t_3$: (chloroquine, inhibits, 2019-ncov).

Following an analysis of the text in COVID-Fact and its factual statements, 
we noticed that not all factual statements have been extracted and transformed into triples in the KG using the above process.
One such example is the triple $t_4$: \textit{(chloroquine, treats, sars-cov-2)}, which we denote as a missing triple, i.e., a fact entailed by the source data but not extracted by the KG construction method.
%We denote such a triple as a \textit{missing triple}. Here, we consider a missing triple to be a fact that is entailed in the source data (e.g., text, table, etc.) but was not extracted by the KG construction method.
%\mourad{This simple definition is fine here, but we should add a formal one later.}
In our running example, $t_4$ can be derived from $f_2$ and is present in the COVID-Fact dataset but was not extracted and is therefore considered a missing triple.

%\fb{a definition of missing triple is missing}
%\textcolor{orange}{
%TODO
%In the extracted triples, some are missing, using manual assessment, we want to find theses missing triples

%To complete $t_1,t_2$, we generate $t_4$
%r,t exist in elem but not h
%}
%Notice that the triple (chloroquine, treats, sars-cov-2), we referred to as $t_4$, is entailed in $f_2$, and explicitly present in the COVID-Fact dataset, but not extracted.  and is thus considered as a missing triple. 

\begin{figure}[!h]
        \centering
        \includegraphics[width=.9\columnwidth]{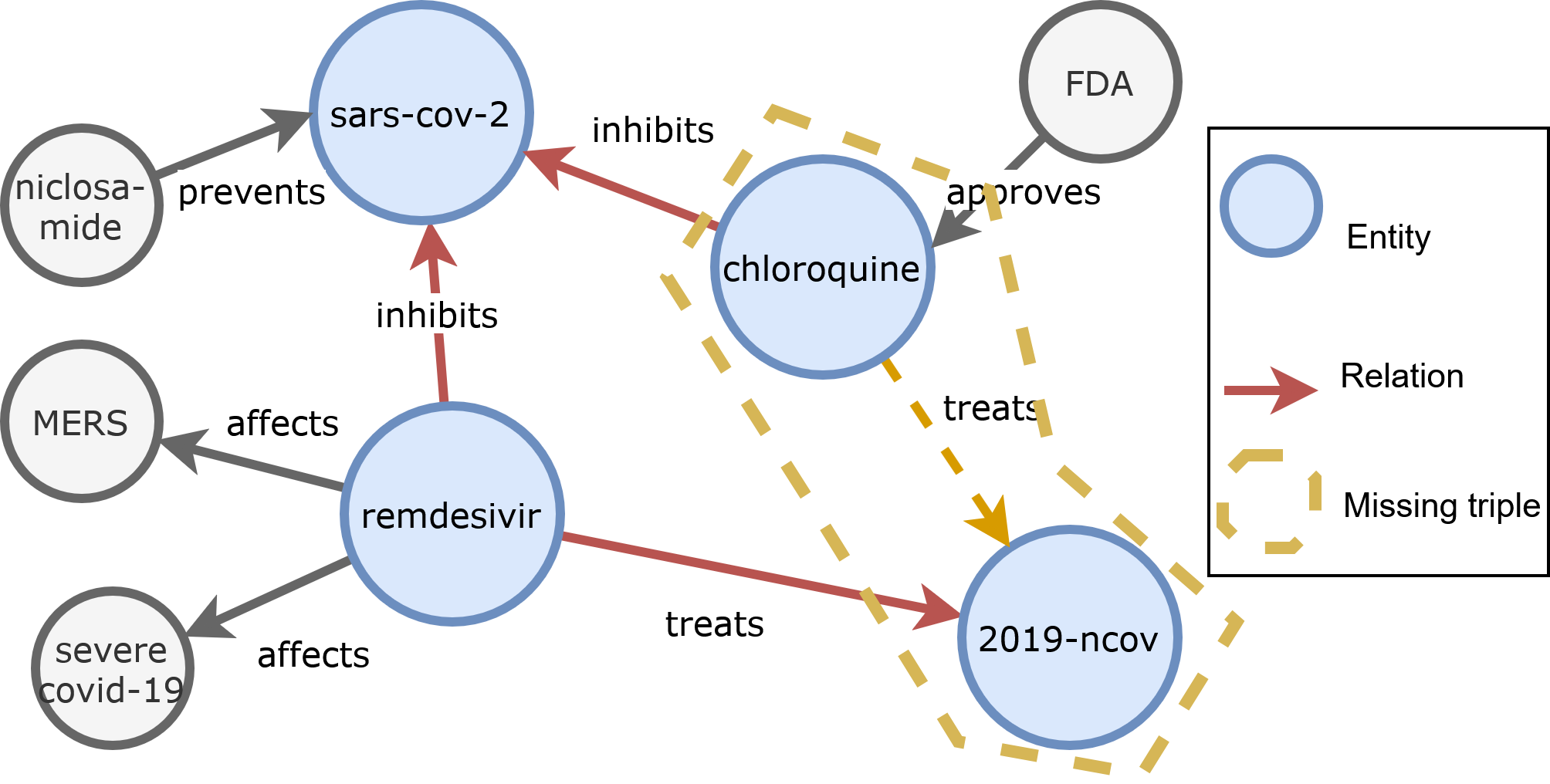}
        \caption{Example of a missing triple in a LLM-generated knowledge graph.}
        \label{fig:missing_triple}
\end{figure}

Figure~\ref{fig:missing_triple} shows the extracted triples and the missing one. We observe that $t_2$ and $t_3$ share the structure \textit{(subject, inhibits, 2019-nCoV)}, suggesting that \textit{remdesivir} and \textit{chloroquine} are semantically related, as they link to the same object through the same relation. A naïve solution to recover missing triples like $t_4$ would be to generate all possible entity–relation combinations, but this yields numerous non-valid or meaningless candidates. Such brute-force search is inefficient and impractical, highlighting the need for a targeted and efficient strategy to generate valid triple candidates while minimizing the search space.
%\textcolor{red}{
While this example illustrates a missing triple caused by incomplete extraction from source data, in practice, the exact causes of missing triples are often unknown, especially in real-world and LLM-generated KGs. Accordingly, evaluating methods designed to recover such triples requires approximations that allow for controlled and reproducible experimentation, as adopted in our experiments.
%}

\noindent{\textbf{Challenges.}} 
The motivating example illustrates three key challenges. 
First, many missing triples are implicitly present in the source text but not extracted, requiring candidate generation methods that can recover such facts efficiently.
Second, most existing approaches rely on external knowledge, which is often unavailable or unreliable. In contrast, our work leverages only the internal structure of the KG, ensuring a controlled and schema-consistent completion process.
Third, given the large number of candidate triples, an initial filtering step is needed to discard non-valid ones and focus evaluation on the most plausible candidates.
%Third, given the potentially large number of candidate triples, it is essential to implement an initial filtering step to eliminate obviously incorrect triples and to limit the assessment step of whether a given triple is missing to a smaller subset that is more likely to be accurate.

\begin{figure}[!h]
    \centering
        \includegraphics[width=0.9\linewidth]{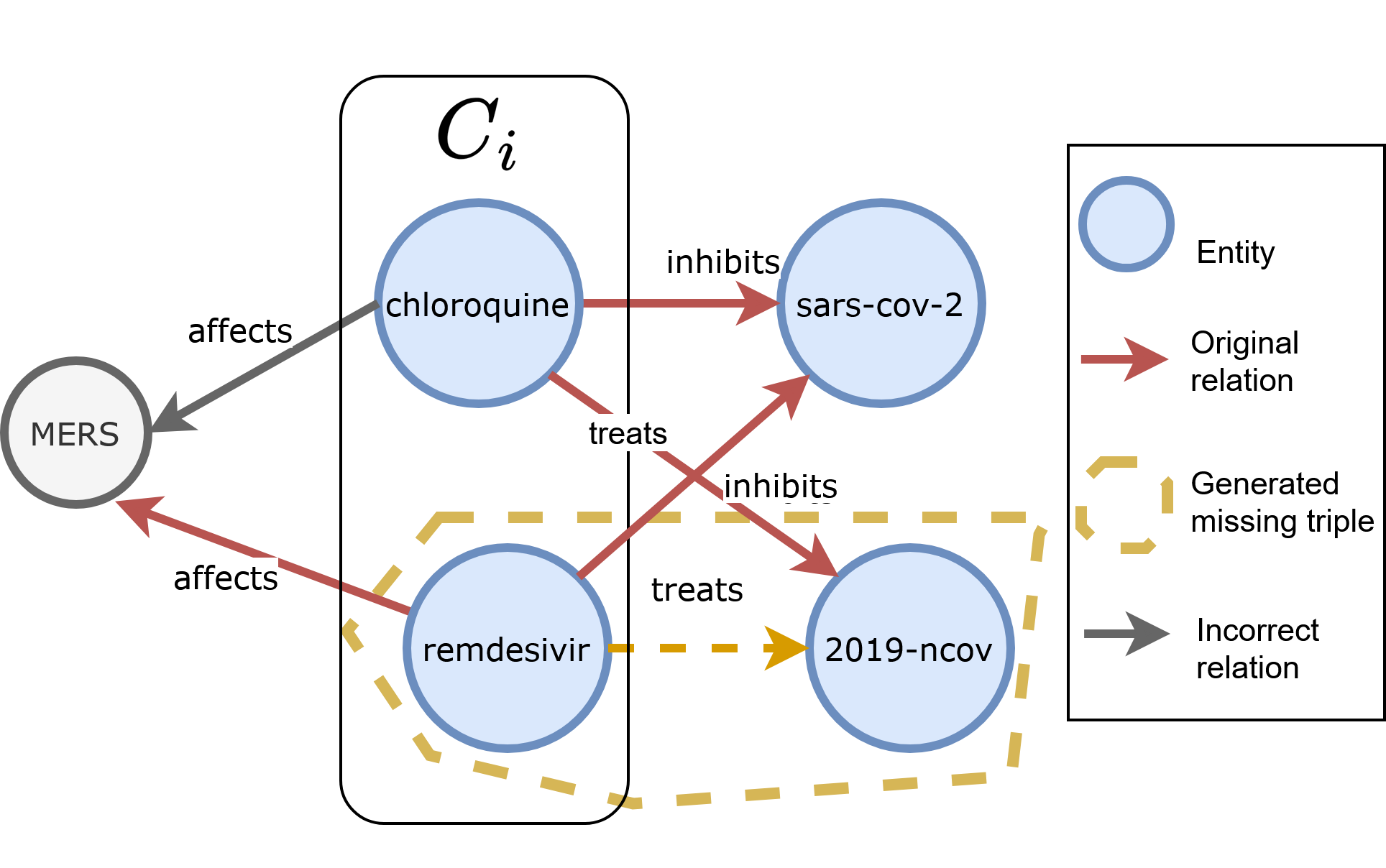}
        \caption{Example of the Clustering-based triple candidate generation.}
        \label{fig:clustering}
\end{figure}

\begin{figure*}
    \centering
    \includegraphics[width=0.85\linewidth]{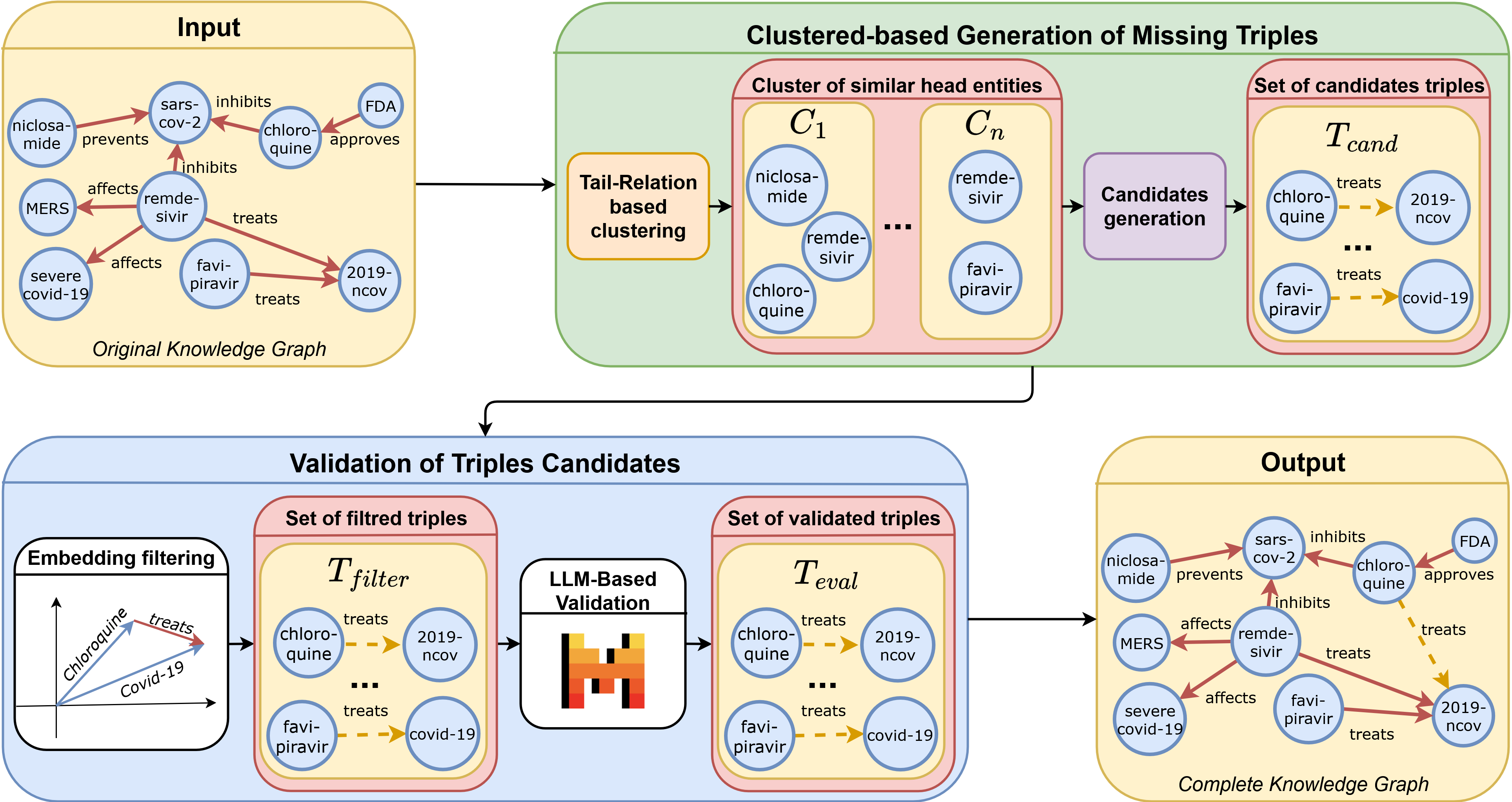}
    \caption{Overview of OMNIA with its two main steps}
    \label{fig:framework-kgc}
\end{figure*}

\noindent{\textbf{Our solution:}}
To address these challenges, we propose  OMNIA\footnote{OMNIA means "everything" in Latin, denotes our aim to close the loop by using LLMs not only to construct but also to complete KGs, while remaining applicable to any KG.}, a two-stage approach for KG completion, as illustrated in Figure~\ref{fig:framework-kgc}:
%mourad{I'm not sure we should limit ourselves to LLM-generated KGs; our method is agnostic to the generation process.}
%tailored to LLM-generated KGs. 
%The name OMNIA reflects our aim of using LLMs not only to create knowledge graphs but also to enrich and complete them. This closes the loop between the generation and completion processes using LLMs. 
%OMNIA involves two main steps depicted in Figure~\ref{fig:framework-kgc}. 

\textit{Clustering-based triple candidate generation}. To address the challenge of missing triples that are implicit in the text but not explicitly extracted during KG construction, OMNIA groups tail entities and relations into clusters according to shared relational patterns. Within each cluster, new triples are inferred by propagating known (relation, tail) pairs to compatible head entities (cf. Figure \ref{fig:clustering}). For instance, \textit{chloroquine} and \textit{remdesivir} are both connected to \textit{SARS-CoV-2} through the relation \textit{inhibits}, thus belong to the same cluster. Knowing that \textit{chloroquine} treats \textit{2019-nCoV} and \textit{remdesivir} affects \textit{MERS}, OMNIA infers candidate triples: \textit{(chloroquine, affects, MERS)} and \textit{(remdesivir,treats, 2019-ncov)}. However, not all generated triples are valid: the second triple is valid, whereas the first one is not valid. This clustering captures valid yet overlooked triples while reducing exhaustive search, improving both coverage and scalability. It leverages the KG’s structural patterns and explicit relation semantics, avoiding dependence on external sources that may be noisy or unavailable, consistent with findings in \cite{yao2025exploringlargelanguagemodels}.
%For instance, \textit{chloroquine} and \textit{remdesivir} are both connected to \textit{SARS-CoV-2} through the relation \textit{inhibits}, thus belong  to the same cluster. Knowing that \textit{chloroquine} treats \textit{2019-nCoV} and \textit{remdesivir} affects \textit{MERS}, OMNIA infers candidate triples: \textit{t1 = (chloroquine, affects, MERS)} and \textit{t2 = (remdesivir,treats, 2019-ncov)}. However, not all generated triples are valid: \textit{t2} is correct, whereas \textit{t1} is incorrect. This clustering captures plausible yet overlooked triples while reducing exhaustive search, improving both coverage and scalability It leverages the KG’s structural patterns and explicit relation semantics, avoiding dependence on external sources that may be noisy or unavailable, consistent with findings in \cite{yao2025exploringlargelanguagemodels}.

%\begin{figure}[ht]
%    \centering
 %   \includegraphics[width=0.9\linewidth]{pb_def/Framework-Problem definition.drawio.png}
  %  \caption{Example of a missing triple in a LLM generated knowledge graph}
   % \label{fig:problem_def}
%\end{figure}

%\begin{figure}[ht]
 %   \centering
  %  \includegraphics[width=0.9\linewidth]{pb_def/Framework-clustering_candidates.drawio.png}
   % \caption{Example of the Clustering-based triple candidate generation}
   % \label{fig:solution}
%\end{figure}

\textit{Validation of missing candidates}. Given the potentially large number of candidate triples, we introduce a two-stage filtering process. A lightweight KG embedding model is first used to eliminate clearly non-valid triples. Then, the most promising candidates are passed to an LLM using different prompting strategies or through a retrieval augmented generation (RAG) system for final validation. This ensures that LLMs are used efficiently while maintaining high precision in the final set of completed triples.

\noindent\textbf{Contributions.} We summarize our contributions as follows.
\begin{itemize}
    \item We introduce OMNIA, a novel approach that generates missing triples with a targeted focus on implicit semantics, often overlooked in LLM-generated KGs. While motivated by these cases, OMNIA remains general and effective across diverse KGs. It leverages only the internal graph structure, using structural clustering of entities and relations with similar patterns to efficiently infer relevant missing facts.
%common in LLM-generated knowledge graphs namely, triples that are entailed in the text but not explicitly extracted. Unlike most existing approaches, our approach leverages only the internal structure of the KG itself, using clustering over head entities with similar relational patterns to infer correct missing facts.\\ 
%While motivated by challenges in LLM-generated KGs, OMNIA is general and effective on existing, non-LLM KGs, as demonstrated in our experiments. Unlike most existing approaches, OMNIA leverages only the internal structure of the KG, using structural clustering over entities and relations with similar relational patterns to infer relevant missing facts.
% rather than all  possible ones.

    \item We propose a two-step validation process that combines structural and textual information from the KG without using external sources. The first stage filters non-valid triples using an embedding model, while the second validates the remaining ones in both structured and natural language scenarios using zero-shot, in-context, and retrieval-augmented prompting to assess semantic validity.  

    \item We conduct an extensive experimental evaluation on multiple KGs, including those generated by LLMs. Our proposed approach outperforms strong baselines on triple classification, with F1-score improvements reaching up to 23 percentage points.% to
% in three of the four datasets used for experiments.  %We experimentally evaluate our approach on several knowledge graphs including LLM generated ones and compare the results to baselines.....  %we provide statistics about each component of OMNIA.
\end{itemize}

The rest of this paper is organized as follows. 
Section~\ref{sec:related} reviews related work. % on KG quality assessment and KG completion techniques, with a focus on recent LLM-based approaches. 
Section~\ref{sec:prob} formalizes the KGC problem. Section~\ref{sec:overview} presents OMNIA with its two main steps. Section~\ref{sec:exp} presents the experimental setup and results. Finally, Section~\ref{sec:conclusion} concludes with future directions.

\begin{comment}
\begin{figure*}[!htbp]
    \centering
    \begin{minipage}[t]{0.4\textwidth}
        \centering
        \includegraphics[width=\linewidth]{pb_def/Framework-Problem definition.png}
        \caption{Example of a missing triple in a LLM-generated knowledge graph.}
        \label{fig:problem_def}
    \end{minipage}
    \hfill
    \begin{minipage}[t]{0.4\textwidth}
        \centering
        \includegraphics[width=\linewidth]{pb_def/Framework-clustering_candidates.png}
        \caption{Example of the Clustering-based triple candidate generation.}
        \label{fig:solution}
    \end{minipage}
\end{figure*}
\end{comment}

%\bibliographystyle{ACM-Reference-Format}
\section{Related Work}
\label{sec:related}
Improving the quality of KGs is crucial for maximizing their usability across various applications~\cite{huaman2021knowledge}. To assess their quality, several dimensions are used, including accuracy, completeness, consistency, conciseness, and relevancy~\cite{zaveri2016quality,WANG2021607}. 
In this paper, we focus on the completeness of KGs.%, focusing on extracting missing triples entailed by the existing triples in the input KG.
%which \textcolor{orange}{address issues about missing triples.}
%or incorrect triples, redundant entries, and erroneous assertions.

\subsection{Knowledge Graph Quality Assessment}
Several studies have proposed frameworks for assessing and refining KGs. Paulheim \cite{paulheim2017knowledge} classified refinement methods by goal (completion vs. error correction) and data source (internal vs. external). Huaman \cite{huaman2021knowledge} introduced a curation framework addressing correctness and completeness through 20 quality dimensions, including duplicate resolution and conflict detection. Wang et al. \cite{wang2021knowledge} proposed a universal framework that evaluates accuracy, consistency, and trustworthiness through the selection and extraction phases. 

\subsection{Knowledge Graph Completion}
As part of KG refinement, KGC tasks include three main tasks \cite{biswas2020entity, shen2022comprehensive}: \textit{link prediction}, which predicts a missing entity or relation within an incomplete triple; \textit{triple classification} which validate the accuracy of a given triple; and \textit{relation prediction} which identifies the appropriate relation between a pair of entities. In some cases, entity classification is also considered, especially when assigning types to entities supports KG enrichment \cite{moon2017learning}. To address these tasks:

\textbf{Traditional embedding based approaches} represent entities and relations in vector spaces to capture graph structure \cite{bordes2013translating, lin2015learning, trouillon2016complex}, but often overlook semantic information. %Triple-based methods, often using structural information like embeddings \cite{bordes2013translating, lin2015learning, trouillon2016complex}, are commonly employed for KG completion. These approaches typically represent entities and relations in a continuous vector space to capture the structural patterns within the knowledge graph.

\textbf{Language model-based approaches} address this limitation by using pretrained models \cite{brown2020language}. For example, KG-BERT \cite{yao2019kg} reformulates triples as textual sequences but struggles with relational disambiguation. StAR \cite{wang2021structure} and SimKGC \cite{wang2022simkgc} improve contextual representations through multitask and contrastive learning, while GPHT \cite{zhang2024start} introduces Triple Set Prediction (TSP) to generate new triples using subgraph embeddings without relying on external text.

%Example of these approaches include KG-BERT \cite{yao2019kg} to score entity-relation-entity sequences as textual inputs for BERT. However, KG-BERT faces two challenges: difficulty of learning relational data and identifying correct answers among similar candidates. 
%To address this, multi-task learning methods that combine relation prediction with relevance ranking and link prediction have been proposed \cite{kim2020multi}.

%Further recent KG completion methods include the structure-augmented text representation (StAR) model \cite{wang2021structure}, which improves link prediction by using a Siamese-style textual encoder for contextualized representations. 
%Similarly, SimKGC \cite{wang2022simkgc} enhances text-based KG completion through efficient contrastive learning with a bi-encoder architecture. GPHT \cite{zhang2024start} addresses KG completion through the Triple Set Prediction (TSP) task by generating new triples without having the head, relation and tail in contrast to other tasks in KGC. GPHT uses the KG's embeddings trained on subgraphs of the KG without relying on external text sources to generate pairs of entities that are likely to be connected by a relation. They then use embeddings trained on the whole KG to predict the relation that most likely to connect the pair of entities.
%\textcolor{orange}{Furthermore, GPHT \cite{zhang2024start}  generate missing triple leveraging KGE.}
%Some method leverage CNN on top of text embedding to predict type \cite{biswas2021contextual}

\textbf{Sequence-to-sequence and training-free approaches} redefine KGC as a generation task. 
KG-S2S~\cite{chen2022knowledge} and 
GenKGC~\cite{xie2022discrimination} improve representation learning with relation-guided demonstrations and entity-aware decoding, while KICGPT~\cite{wei2024kicgpt} enhances efficiency through in-context learning without fine-tuning. %\textcolor{red}{
KGT5 \cite{saxena-etal-2022-sequence} uses an encoder–decoder transformer with auto-regressive decoding for scalable KG link prediction.
%}

%\textbf{Sequence-to-Sequence and Training-Free Approaches.} Sequence-to-sequence pre-trained language model (PLM) methods redefine KG completion as a generation task. Examples include KG-S2S \cite{chen2022knowledge} and GenKGC \cite{xie2022discrimination}. The latter improves representation learning with relation-guided demonstrations and entity-aware hierarchical decoding. Recent training-free approaches like KICGPT \cite{wei2024kicgpt} further address the limitations of fine-tuning, offering greater efficiency in KG completion.

\textbf{Large language model approaches} further integrate semantic reasoning with structural knowledge. KG-LLM \cite{shu2024knowledge} transforms triples into natural language to improve reasoning, and MuKDC \cite{li2024llm} generates potential missing data filtered by embeddings. Hybrid methods such as KoPA \cite{zhang2024makinglargelanguagemodels} and MPIKGC \cite{xu2024multi} combine structural embeddings or LLM-generated descriptions to enhance inference.
%\textbf{Large Language Model Approaches.} Recent approaches have leveraged LLMs to address Knowledge Graph Completion tasks. LLMs have shown the potential to improve link prediction performance on standard benchmarks. %For instance, KG-LLM \cite{shu2024knowledge} transforms Knowledge Graph data into natural language to enable LLM-based reasoning; MuKDC \cite{li2024llm} use LLM to generate potential missing data and employs an embedding-based method to filter out incorrect ones and KoPA \cite{zhang2024makinglargelanguagemodels} combine integrates structural information from the graph with LLM capabilities to improve it result.
%LLMs can be used to enhance existing knowledge graph completion methods, such as MPIKGC \cite{xu2024multi}, which employs an LLM to generate descriptions for entities and/or relations. several works have further explored how to more effectively integrate structural and semantic reasoning within LLMs. Approaches such as KICGPT \cite{wei2024kicgpt} leverage in-context learning to encode structural knowledge without additional fine-tuning. KG-LLM~\cite{yao2025exploringlargelanguagemodels} transforms triples into text sequences to enhance reasoning, and other approaches combine graph transformers with LLMs \cite{luo2025gltwjointimprovedgraph} or incorporate structural embeddings as virtual knowledge tokens to improve connectivity and inference accuracy~\cite{zhang2024makinglargelanguagemodels}

\subsection{OMNIA's take on the Problem}
Unlike traditional embedding-based methods that rely on structural patterns \cite{bordes2013translating,yang2015embeddingentitiesrelationslearning,trouillon2016complex}, OMNIA combines both the graph’s structure and the semantic capabilities of LLMs to capture missing triples, particularly in cases where triples are implicitly entailed. Compared to language model-based methods \cite{yao2019kg,wang2021structure,wang2022simkgc}, OMNIA does not rely on external textual corpora. Instead, like GPHT \cite{zhang2024start}, it generates candidates directly from the KG. However, OMNIA specifically targets challenges in LLM-generated KGs, in particular incomplete extraction, which GPHT does not consider. Finally, unlike approaches that use LLMs for end-to-end prediction, OMNIA applies LLMs in a targeted validation step, after filtering structurally unlikely triples, which makes it both efficient and adaptable to real-world KGs with varying characteristics. 

%\textcolor{red}{
Overall, OMNIA adopts a system-level design that integrates structural and semantic reasoning for closed-world KG completion, relying solely on the input KG. In contrast to approaches that integrate semantic reasoning directly into learned representations or hybrid embeddings, OMNIA treats semantic validation as a separate and explicit step, enabling efficient and controlled use of LLMs in closed-world  settings.
%}

%Overall, our approach can be seen as a synthesis of existing strategies, combining novel candidate generation, effective filtering, and robust LLM-based validation. \textcolor{red}{improve this last sentence to consider [R7O1]}
%\textcolor{orange}{We propose OMNIA, a Knowledge Graph Completion framework. A central idea to OMNIA is to use both structural and textual information of the KG, which would lead to the best results.
%In order to achieve the best result, our method OMNIA uses both structural and textual information of the KG. We first generate potential missing data, for this generation part we use the structural information of the KG similar to GPHT~\cite{zhang2024start}. Other methods such as MuKDC~\cite{li2024llm} leverage LLM instead. Then, we assess the validity of these generated triples in two steps: we first filter out some incorrect data using a traditional embedding method, namely TransE \cite{bordes2013translating}, and then classify each triples using an LLM. Our proposed approach can be applied to any KG, either generated using LLMs or using other methods.
%Unlike previous method, our method also work on LLM-generated KG. Our method can be seen as a synthesis of existing methods with new creative ways to generate candidate missing triples, filter them out, and finally validate them using LLMs\mourad{What do you think about this last sentence!?}.}
%\textcolor{orange}{Frederic: The sentence is very elegant but is our method really a synthesis of existing method ?}
\section{Problem Definition}
\label{sec:prob}
% define KG 
%\textit{\textbf{Definition 1.}  

Let us consider a \textit{Knowledge Graph}, \(\mathcal{G} = (\mathcal{E}, \mathcal{R}, \mathcal{T})\), where \(\mathcal{E}\) is a set of entities, \(\mathcal{R}\) is a set of relations, and \(\mathcal{T}\) is a set of triples. Each triple \((h, r, t) \in \mathcal{T}\) consists of a \textit{head entity} \(h \in \mathcal{E}\), a relation \(r \in \mathcal{R}\), and a \textit{tail entity} \(t \in \mathcal{E}\). An example of a triple is \textit{(Probiotics, Prevents, Covid-19 infection)}. %In this work, we assume that \(\mathcal{G}\) is automatically constructed from a source corpus using a Large Language Model (LLM).
%\textbf{Definition 2.} \textbf{\textit{Knowledge Graph Completion}} 

\textit{Knowledge Graph Completion (KGC)} is the task of recovering triples that are missing from a KG. It may either leverage external sources, such as text corpora, structured databases, or rely solely on the internal structure and content of the graph. 
This involves two common assumptions \cite{shi2018open}: the open-world assumption where new entities or relations not present in the original graph may be introduced, and the closed-world assumption, which limits the completion to combinations of existing entities and relations in the graph. 
%To address the KGC problem, several task formulations have been proposed \cite{shen2022comprehensive}, including \textit{link prediction}, which predicts a missing entity or relation within an incomplete triple; \textit{triple classification} which assesses the plausibility of a given triple; and \textit{relation prediction} which identifies the appropriate relation between a pair of entities. In some cases, entity classification is also considered, especially when assigning types to entities supports Knowledge Graph enrichment \cite{moon2017learning}.

In this work, we adopt a closed-world setting and focus on KGC operating solely on the KG itself. Unlike link prediction or relation prediction, which assume partially observed triples, our objective is to generate entirely new triples that are \textit{entailed} by the facts but not explicitly present in the graph. 
In addition, while triple classification typically involves binary plausibility prediction over predefined candidates, our KGC problem extends this by including a dedicated LLM-based validation step to evaluate the generated triples in context, as we present in Section \ref{sec:LLMbased}. 
%Although our KGC problem is applicable to any KG, it particularly considers challenges posed by KGs constructed using LLMs. 
%In particular, we focus on a type of missing triples that often arises in such graphs, as illustrated in the motivating example above. 

\textit{Missing triples} are triples that are not present in the KG, yet  are logically entailed by the source text used to construct it. Importantly, the entities and relations involved in such triples are already included in the KG, only the complete statement connecting them is missing. 
%\mourad{Why are we adding the following sentence?}%This type of incompleteness arises not from the unavailability of external knowledge, but from the limitations of extraction models. 
Formally, we define a missing triple in our setting as a triple \( (h, r, t) \notin \mathcal{T} \) such that \(h, t \in \mathcal{E}\), \(r \in \mathcal{R}\), and \((h, r, t)\) is entailed by the source corpus used to generate \(\mathcal{G}\). While such missing triples are particularly common in KGs constructed from text using LLMs, our formulation and proposed approach apply to any KG where similar types of incompleteness may arise.
\section{Overview of OMNIA}
\label{sec:overview}

We introduce OMNIA, a two-stage approach designed to enhance the completeness of KGs, particularly those generated by LLMs. As depicted in Figure~\ref{fig:framework-kgc}, the first stage generates missing triple candidates through a cluster-based generation process and the second validates these generated triples through a two-step process consisting of an embedding-based filtering followed by LLM-based validation. By combining these two stages, OMNIA bridges the gap between structural and semantic reasoning in KG completion. Its originality lies in relying on the internal structure and semantics of the KG rather than external textual sources, while still integrating LLMs for selective semantic validation.

OMNIA is designed as a system-level approach, rather than as a new embedding or learning architecture. This design enforces a clear separation between structural inference and LLM-based semantic validation, allowing each stage to focus on a well-defined role and enabling efficient and controlled use of LLMs in closed-world KG completion settings.

To illustrate how our proposed two-stage design operates in practice, we consider the case of the missing triple $t_4$, form the running example (cf. Section \ref{sec:intro} ). In the first stage, entities sharing similar relational contexts (e.g., \textit{chloroquine} and \textit{remdesivir}) are grouped into the same cluster. The underlying intuition is that entities connected through similar relations tend to exhibit a similar behavior, enabling OMNIA to uncover implicit triples like $t_4$. In the second stage, the inferred triples are filtered and validated, retaining only semantically valid triples, such as $t_4$. 
In the rest of the section, we detail each stage of OMNIA and then analyse its complexity.

\subsection{Clustered-based Generation of Missing Triples}
\label{sec:sub_cluster}
This stage aims to infer missing triples by leveraging the internal structural patterns of the KG. OMNIA groups entities and relations into clusters based on their relational similarity, computed from the graph’s connectivity patterns rather than external textual features. 
This clustering step allows the efficient generation of a high-coverage set of candidate triples by exploiting recurring structural patterns in the graph, while deferring semantic reasoning to the next stage of the approach.

Although this is motivated by patterns frequently observed in LLM-generated KGs, as introduced in the motivating example, the same mechanism generalizes to any KG and can be adapted to different assumptions of missing knowledge beyond implicit semantics, as demonstrated in our experiments. %\textcolor{red}{Semantic reasoning is handled explicitly in the subsequent validation stage using an LLM.} 
This stage consists of two main steps presented below.%Within each cluster, relational patterns are propagated across compatible entities to generate candidate triples that are consistent with the existing KG schema.
%\textcolor{blue}{In our previous work~\cite{ezzabady2024towards}, we generated a KG using an LLM and observed that semantically similar entities often share similar relations with the same tail entities, a pattern frequently observed in LLM-generated KGs. For example, entities such as \textit{COVID-19} and \textit{SARS-CoV-2} may both appear in relations such as \textit{has symptom} or \textit{treated by}. Based on this observation, we propose to cluster similar entities and infer new triples within each cluster, under the assumption that entities grouped together are likely to share similar relations. This clustering-based strategy supports KG completion by generating missing but plausible facts. To ensure that the generated triples are coherent with the existing KG and represent genuinely missing knowledge, the generation process relies solely on information already contained within the KG. The generation of missing triple candidates consists of two main steps: (i)~clustering semantically similar entities and (ii)~systematically constructing candidate triples. These two steps are presented below.}

\begin{algorithm}
\caption{Clustering of head entities}
\label{alg:ent_clustering}
\begin{algorithmic}[1]
\Require A knowledge graph $\mathcal{G} = \{(h_i, r_i, t_i)\}_{i=1}^n$
\Ensure A set of clusters $\mathcal{C} = \{\mathcal{C}_1, \mathcal{C}_2, \dots\}$

\State Initialize an empty dictionary: $\mathcal{C}\gets$ \{\}

\ForAll{$(h_i, r_i, t_i) \in \mathcal{G}$}
    \State $k \gets (r_i, t_i)$ \Comment{Tail-relation pair as cluster key}
    \If{$k \in$ $\mathcal{C}$}
        \State $\mathcal{C}$[$k$] $\gets$ $\mathcal{C}$[$k$] $\cup \{h_i\}$
    \Else
        \State $\mathcal{C}$[$k$] $\gets \{h_i\}$
    \EndIf
\EndFor
\State \Return $\mathcal{C}$
\end{algorithmic}
\end{algorithm}

\subsubsection{Entity Clustering}
%\noindent\textbf{Entity Clustering}
Let us consider $\mathcal{G}$, a KG such as:
\[
\mathcal{G} = \{ (h_1, r_1, t_1), (h_2, r_2, t_2), \dots, (h_n, r_n, t_n) \}
\]
where $(h_i, r_i,t_i)$ represents a triple with $h_i$ representing the head entity, $r_i$ the relation and $t_i$ the tail entity.

The clustering concept in OMNIA is grounded in the intuition that entities appearing in similar structural roles within a KG, namely entities occurring in comparable relational contexts, tend to encode related semantic properties.

%The clustering concept in OMNIA is based on the intuition that entities appearing in similar structural roles in a knowledge graph, such as appearing in comparable relational contexts, tend to encode semantically related meanings.
%Based on the manual quality assessment conducted in our previous work \cite{ezzabady2024towards}, we assume that head entities sharing a relation with the same tail entity are likely to be semantically similar.
To this end, OMNIA groups structurally similar head entities into clusters to generate candidate missing triples.

Algorithm~\ref{alg:ent_clustering} outlines this process.
%Our goal is then to cluster these similar head entities in order to generate candidate missing triples. Algorithm~\ref{alg:ent_clustering} outlines the clustering of these head entities.
First, for each triple in $\mathcal{G}$, we extract the relation-tail pair $(r_i,t_i)$ (Line 3 in Algorithm \ref{alg:ent_clustering}). If a cluster corresponding to this $(r_i,t_i)$ pair already exists, the head entity $h_i$ is added to that cluster (Lines 4-5). Otherwise, a new cluster is created for the pair (Lines 6-7). The algorithm returns the set of clusters, denoted by $C$, each corresponding to a distinct relation–tail pair observed in $\mathcal{G}$.
%If a cluster exists for the $(r_i,t_i)$ pair, the head entity, $h_i$ is added to that cluster (Lines 4-5). Otherwise, a new cluster is created for the pair (Lines 6-7). The algorithm returns the set of clusters $C$, containing all clusters corresponding to the relation-tail pairs in $\mathcal{G}$.

\begin{figure}
    \centering
    \includegraphics[width=0.75\linewidth]{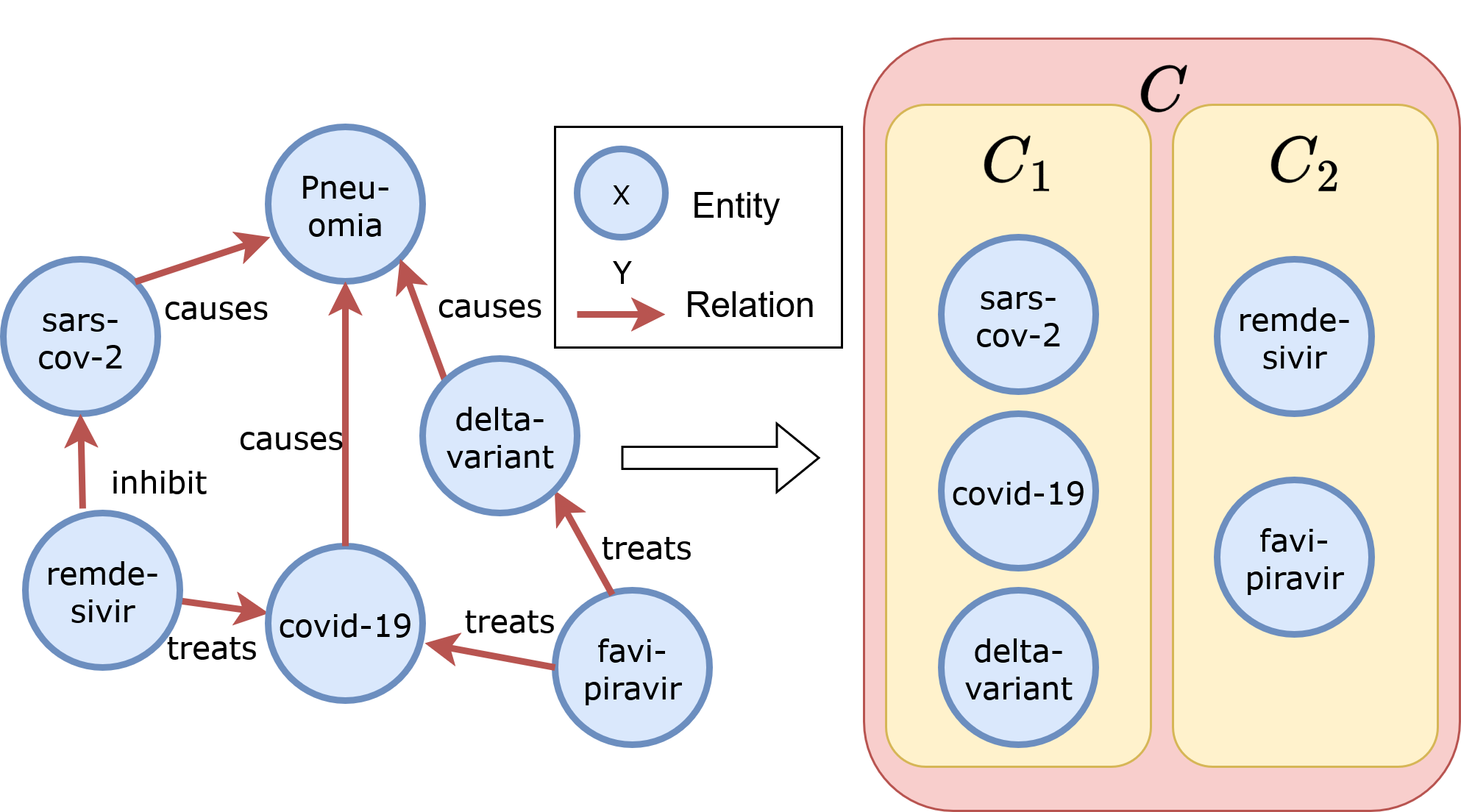}
    \caption{Example of Entity Clustering}
    \label{fig:ent_cluster_example}
\end{figure}

For instance, by applying our clustering to the following triples:\\
%$\mathcal{G}$ = \
\{
(\textit{sars-cov-2, causes, pneumonia)},
\textit{(covid-19, causes, pneumonia)},
\textit{(delta-variant, causes, pneumonia)},
\textit{(remdesivir, treats, covid-19)},
\textit{(remdesivir, inhibit, sars-cov-2)},
\textit{(favipiravir, treats, covid-19)},
\textit{(favipiravir, treats, delta-variant)}\} \\
\noindent We obtain the following clusters:\\ $\mathcal{C}_1$ = \{\textit{sars-cov-2, covid-19, delta-variant}\}, where all entities share the relation-tail pair \textit{(causes, pneumonia)}; \\
and $\mathcal{C}_2$ = \{\textit{remdesivir, favipiravir}\}, where both entities share the relation-tail pair \textit{(treats, covid-19)}. Figure \ref{fig:ent_cluster_example} illustrates this example.
We note that the same head entity may belong to multiple clusters if it appears in several triples sharing the same relation–tail pair.
%\textcolor{red}{Add for the overlap with an example, [R2O5]}

\subsubsection{Candidate Triples Generation}
\label{subsection:cand_gen}
Let us consider $C$ such as:  $C = \{C_1,C_2,\dots, C_K\}$, the set of clusters obtained from the previous clustering step. These clusters are leveraged to systematically generate candidate missing triples, as depicted in Algorithm~\ref{alg:gen_miss_candidate}.
For each cluster \( C_i \in C \), we identify all head entities \( h_i \in C_i \) (Line 3). 
For each of these entities, we collect the relation–tail pairs \( (r_i, t_i) \) with which they are associated (Line 4). Candidate triples are generated by combining each head entity within \( C_i \) with every relation–tail pair extracted from the same cluster (Line 7). This process results in new triples of the form \( (h_j, r_i, t_i) \), which are treated as candidates for missing triples in the KG.
%plausible candidates for missing triples in the KG.

%We then extract all relation-tail pairs \( (r_i, t_i) \) associated with these head entities (Line 4). The candidate triples are generated by systematically combining each head entity in \( C_i \) with every relation-tail pair from the same cluster (Line 7). This results in new triples of the form \( (h_j, r_i, t_i) \), which serve as potential missing triple in the knowledge graph. 
These candidates are subsequently validated in the next stage of OMNIA, as described below. We note that our generation process may lead to fewer candidates in sparse KGs, as it relies on structural patterns present in the KG. We discuss this behavior in our experimental analysis (Section \ref{sec:exp}-D1) and consider extensions to handle sparsity as part of future work.

%Let us consider $C$, a set of clusters, such as:
%\[C = \{C_1,C_2,\dots, C_K\}\]

%\noindent we leverage $C$ to generate the missing triple candidates, as depicted in Algorithm~\ref{alg:gen_miss_candidate}.
\begin{comment}
Once the clusters of semantically similar entities are built, we generate our triple candidate with the algorithm below.    
 \textcolor{red}{add more formalization and refer to line in the algo}
~\ref{alg:gen_miss_candidate}.
\end{comment}
%For each cluster \( C_i \in C \), we identify all head entities \( h_i \in C_i \) (Line 3 in Algorithm~\ref{alg:gen_miss_candidate}). Then, 
%For each cluster $C_i$, we include all triples whose head entity is present in $C_i$, as determined by the prior clustering step.

%For each cluster $C_i$ in $C$, we first extract every head entities,$h_i$ (Line 3 in Algorithm~\ref{alg:gen_miss_candidate}).
%We also extract every relation-tail pair $(h_i,r_i)$ associated with any head entity within $C_i$ (Line 4 in Algorithm~\ref{alg:gen_miss_candidate}).
%We then systematically recombine these pairs with every head entity in the same cluster to generate missing triple candidates (i.e., triples that may correspond to missing facts in the KG), as shown in Line 7 of Algorithm~\ref{alg:gen_miss_candidate}.

\begin{algorithm}[htbp]
\caption{Candidates generation}
\label{alg:gen_miss_candidate}

\begin{algorithmic}[1]
\Require A set of clusters $\mathcal{C} = \{C_1, C_2, \dots, C_K\}$ \Comment{Each $C_k$ contains triples $(h_i, r_i, t_i)$}
\Ensure A set of candidate triples $\mathcal{T}_{\text{cand}}$

\State $\mathcal{T}_{\text{cand}} \gets \emptyset$
\ForAll{$C_k \in \mathcal{C}$}
    \State $H_k \gets \{ h_i \mid (h_i, r_i, t_i) \in C_k \}$
    \State $P_k \gets \{ (r_i, t_i) \mid (h_i, r_i, t_i) \in C_k \}$
    \ForAll{$h_i \in H_k$}
        \ForAll{$(r_i, t_i) \in P_k$}
            \State Add $(h_i, r_i, t_i)$ to $\mathcal{T}_{\text{cand}}$
        \EndFor
    \EndFor
\EndFor
\State \Return $\mathcal{T}_{\text{cand}}$
\end{algorithmic}
\end{algorithm}

\subsection{Validation of Candidate Triples}
\label{sec:validation}
%Following the generation of missing candidates, we acknowledge that 
Not all generated triples are necessarily semantically valid. 
To address this issue, this stage first removes structurally unlikely triples using KG embeddings, and then relies on LLMs to assess the semantic validity of the remaining ones. %filters missing candidate triples using KG embeddings and then uses LLMs to validate these filtered candidates.

%The procedure first applies an embedding-based filtering step to remove structurally unlikely candidates, and subsequently relies on LLMs for semantic validation of the remaining triples.

\subsubsection{Embedding-Based Filtering of Candidate Triples}
\label{sec:sub_method_emb}
%\noindent\textbf{Filtering using Knowledge Graph Embedding.}

To filter the generated candidates, OMNIA employs TransE~\cite{bordes2013translating}, a model trained on the original KG to capture latent representations of entities and relations. For each candidate triple $(h, r, t)$, a %distance 
score is computed in the embedding space, where lower values indicate a higher likelihood that the triple is valid.
%the distance between embeddings is computed, where a lower value indicates a higher likelihood that the triple is valid.

For example, given a valid triple \(\mathcal{T}_1  = (h_1,r_1,t_1)\) and a non-valid triple \(\mathcal{T}_2 = (h_1,r_1,t_2)\), the sum of the head entity embedding \(h_1\) and the relation embedding \(r_1\) is expected to be equal or close to the embedding of the true tail entity \(t_1\), and far from that of the non-valid tail entity \(t_2\). Candidate triples whose distance score falls below a threshold $\tau$ are retained, while the others are discarded. This step acts as an efficient pruning mechanism, substantially reducing the candidate space before the subsequent LLM-based validation stage.

We emphasize that TransE serves as a structural validity filter to prune unlikely candidates triples. While its embeddings reflect the observed KG, the model does not explicitly encode assumptions about the distribution of missing triples across relations or entities, relying only on geometric consistency in the embedding space. In other words, we frame TransE as a tool, not a probabilistic model.%does not rely on assumptions about the distribution of missing triples across relations or entities 

%\textcolor{blue}{Salima: In order to block theoretical objections, I would suggest to change the wording as follows: We emphasize that TransE is employed solely as a structural plausibility filter for candidate triples. While its embeddings reflect the observed training data, the model does not explicitly encode assumptions about the distribution of missing triples across relations or entities, relying only on geometric consistency in the embedding space. In other words, we frame TransE as a tool, not a probabilistic model.}

%For example, for a correct triple \(\mathcal{T}_1  = (h_1,r_1,t_1)\) and an incorrect triple \(\mathcal{T}_2 = (h_1,r_1,t_2)\), the sum of the head entity embedding, \(h_1\) and relation embedding, \(r_1\), will be equal or close to the correct triple tail embedding, \(t_1\) but far from the incorrect triple tail embedding, \(t_2\).
%Triples whose distance falls below a threshold $\tau$ are retained, while others are discarded. This step serves as an efficient pruning mechanism, reducing the candidate space before LLM validation.\textcolor{red}{UPDATE [R2O4]}
%We thus only select candidates that have an embedding distance that is below a certain threshold. This threshold should allow us to reduce the number of candidates, but without dismissing the correct candidates.

\subsubsection{LLM-Based Validation of Candidate Triples}
\label{sec:LLMbased}
%\noindent \textbf{Triple Correctness checking using LLMs.}
To check the semantic validity of candidate triples, we employ an LLM as a semantic judge (Algorithm \ref{alg:llm_eval}). 
Based on our problem formulation, which focuses on identifying and validating plausible missing triples, this validation step  assesses the semantic validity of generated candidate triples given the relational structure of the KG. To address this validation objective, the first design decision concerns the format in which candidate triples are presented to the LLM. Specifically, we consider:
%The first design decision in our evaluation framework relates to format in which the information is presented to the LLM. 
%
%We propose two options:
(i)~triple-based validation, where candidates are evaluated in their original structured form, preserving the relational semantics of the KG and facilitating reasoning over entities and relations; and 
(ii)~sentence-based validation, where each triple is transformed into a natural language statement using LLM, allowing the model to leverage its language understanding capabilities to capture implicit semantic information that may not be explicitly encoded in the graph.

\begin{figure*}[!ht]
    \centering
    \includegraphics[width=.7\linewidth]{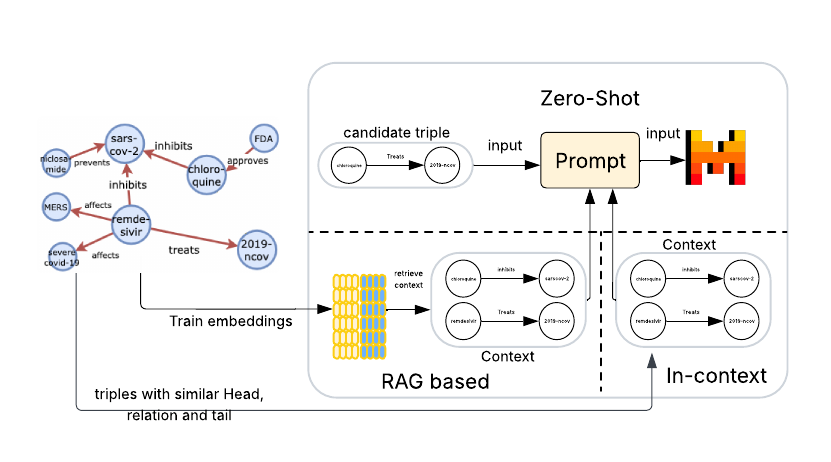}
    \caption{Triple-based scenario with its three prompting cases for LLM-based validation}
    \label{fig:candidates evaluation}
\end{figure*}

\begin{algorithm}[h]
    \caption{LLM evaluation}
    \begin{algorithmic}[1]
    \Require A set of triple candidate $\mathcal{T}_{cand}$
    \Ensure A set of evaluated triples candidate $\mathcal{T}_{eval}$
    \For{each triple ${t} \in \mathcal{T}_{cand}$}
        \State Ask the LLM  to evaluate the candidate
        \If{$t$ is considered true}
            \State $\mathcal{T}_{eval} \leftarrow t$ 
        \EndIf
    \EndFor
    \State Return $\mathcal{T}_{eval}$
    \end{algorithmic}
    \label{alg:llm_eval}
\end{algorithm}
\begin{comment}
Unlike the other parts of the framework that use structural information about KG, this part use the textual information of the KG.
\end{comment}

The second design decision concerns the prompting strategy used during validation, which determines what additional contextual information needs to be provided to the LLM when assessing a candidate triple. In particular, we consider:
%The second design decision is how to interact with or prompt, the LLM. For this, we propose three methods based on the level of contextual information being provided to the LLM:
\begin{itemize}
 %\fb{zero shot prompting}
    \item Zero-shot prompting, the LLM is prompted solely with the candidate triple or sentence.
    This method does not use any form of context nor example and evaluate the triple or sentence as-is.
    This would show the performance of the plain LLM without additional information.
    \item Few-shot prompting, the LLM is provided with a small set of related triples that share a common head, relation, or tail with the candidate triple. These triples serve as contextual examples for the evaluation. 
    This context set includes at least one triple that shares each component (head, relation, or tail) with the evaluated triple. By exposing the LLM to related triples, it gains additional context about the candidate through its "relatives", which helps it provide a more accurate validation.
    %The idea is that by giving information about related triples, the LLM would have more context about the task at hand by learning more about the candidate triple through its "relatives" and hence provide a better answer.
    %know more about each element and will be able to provide better answer.
    \item Retrieval-Augmented Generation (RAG), the LLM receives a top-$k$ set of semantically similar triples retrieved from the embedding space to inform its decision.
    RAG is a widely-used framework in generative AI that combines the strengths of information retrieval systems with the capabilities of LLMs. It constrains the LLM to respond based on retrieved evidence.%It aims to constrain the LLM to respond to a prompt based solely on the information extracted from a curated database (usually a vector database).
    By adding semantically similar triples to the a prompt asking about the validity of the candidate triple, the LLM can make a more informed decision using a more relevant context than few-shot prompting.%the LLM would make a much "better informed" decision. This method would give more relevant triples (forming the context for the final prompt) than the above few-shot prompting.
\end{itemize}

We also tested several prompt templates and iteratively refined them. Figure \ref{fig:candidates evaluation} depicts each case for the triple-based scenario. For the sentence-based scenario, we use an LLM to transform a triple and the corresponding context into a sentence. This requires explicit prompt constraints, as the LLM may otherwise respond by judging the validity of the triple instead of performing the transformation. Therefore, for evaluation, we explicitly instruct the LLM to only transform the triple into a sentence and avoid any negative or evaluative formulation.
%This turns out to be a bit challenging since the LLM would sometimes respond by stating that the triple is incorrect instead of just doing the transformation.
%We have found that if we don't explicitly indicate to only transform triple to sentence and not correct it, it would sometimes transform a triple into it negative form because it knows that the presented information is incorrect.
%It would make the LLM evaluate the wrong sentence.
This is why for the evaluation, we need to explicitly indicate to the LLM to only transform the triple into a sentence and not use any negative format.

%table input>output for each scenario
\begin{comment}
\begin{figure*}
    \centering
    \includegraphics[width=.7\linewidth]{method/smaller.pdf}
    \caption{Triple-based scenario with its three prompting cases for LLM-based validation}
    \label{fig:candidates evaluation}
\end{figure*}
\end{comment}

\subsection{Complexity analysis}
\label{subsection:complexity}
%This section analyzes the 
%\textcolor{blue} {Salima: I would precise time complexity because we are not talking about space complexity} computational complexity of our proposed OMNIA approach across its two stages. 
The clustering step starts by creating an empty dictionary in $O(1)$ time and iterates once over all triples in the KG $G$. For each triple, it builds a key from the relation–tail pair and performs constant-time dictionary operations to update clusters. 
Since each triple is processed exactly once, the total time complexity of this step is $O(n)$, where $n$ is the number of triples in the KG.
In the candidate generation step, each cluster $C_k$ produces $O(m_k \times p_k)$ combinations, where $m_k$ and $p_k$ denote the number of head entities and relation--tail pairs within the cluster, respectively. The overall cost is therefore $O(\sum_k m_k \times p_k)$.
In the worst case, a single cluster contains all $n$ triples, yielding at most $n$ distinct head entities and n distinct relation–tail pairs. Consequently, the candidate generation step admits a worst-case upper bound of $O$($n^2$).
%\textcolor{blue}{Salima : I would change the wording-  Consequently, the candidate generation step admits a worst-case upper bound of O($n^2$).}
%In the worst case scenario, every entity is connected to every entity leading to a $O(n^3)$ complexity.
The embedding-based filtering is linear, with complexity $O(n + n')$, where $n'$ is the number of generated candidates.
Finally, the LLM-based validation evaluates each remaining candidate once, resulting in $O(n')$.
Overall, the total complexity of OMNIA is $O(n + \sum_k m_k \times p_k + n')$, which is dominated by the candidate generation step.

\section{Experimental Evaluation}
\label{sec:exp}
%\textcolor{red}{complete the results of PLM based in the table}
We conducted comprehensive experiments on OMNIA and addressed the following key questions:
\begin{itemize}
    \item (Q1) How efficient and effective is OMNIA’s candidate generation in identifying relevant missing triples compared to an exhaustive strategy?

    \item (Q2) How effective is our evaluation model compared to the SOTA baselines from three different classes of methods?
%/(Q5) How does our method performs on dense and sparse KG?\
    
    \item (Q3) What is the effect of using sentence-based versus triple-based prompting in the LLM-based evaluation?

    \item (Q4) What are the key parameters that impact the performance of OMNIA’s prompting strategies, particularly in RAG?
\end{itemize}

%\subsection{Experimental setting}

\subsection{Datasets and Evaluation Metrics}
\label{subsection:evalMetr}
We conducted our experiments on the commonly used datasets for KG completion tasks and KG embeddings evaluation~\cite{shu2024knowledge,yao2025exploringlargelanguagemodels,wei2024kicgpt} in addition to a KG generated using an LLM as described in our previous work~\cite{10.1007/978-3-031-81221-7_23} to assess performance in LLM-generated KGs.

\begin{itemize}
    \item FB15K-237 \cite{toutanova2015observed} is a refined version of the FB15K~\cite{bordes2013translating} dataset which is originally a subset of Freebase. 
    This version is specially designed to avoid the data leakage problem discovered with the FB15K. The original FB15k contained many inverse relation pairs, which led to inflated performance metrics in KG completion tasks. FB15k-237 resolves this issue by removing such inverse relation pairs. The dataset includes common sense knowledge as well as complex relational facts across a variety of topics.
   
    \item CoDEx-M \cite{safavi2020codex} is the medium-scale subset of the CoDEx (Contextualized Decoding for KG Completion) benchmark. It is derived from Wikidata \cite{10.1007/978-3-031-81221-7_23} and includes natural names for entities and relations, making it well-suited for language-grounded KG tasks. The dataset is designed to better capture semantic diversity and support contextual reasoning.% in KG completion.
   
    \item WN18RR\cite{dettmers2018convolutional} is an improved version of WN18~\cite{bordes2013translating}, which is a subset of the well-known WordNet dataset. %WordNet is a large lexical database of English nouns, verbs, adjectives, and adverbs connected through conceptual-semantic and lexical relations. 
    %While WN18 was originally extracted from WordNet, it was found to contain a large number of test triples that were simply inverse of those in the training set \textcolor{blue}{provided by the original dataset} \mourad{Which training set are you referring to?}, resulting in data leakage and inflated evaluation scores. WN18RR addresses this issue by removing such inverse relations from the dataset.
The original WN18 dataset was derived from WordNet but was later found to include a large number of test triples that were simple inverses of triples present in its training split (i.e., the training portion of the WN18 benchmark dataset).

    \item Socio-Economic is a KG generated using LLMs~\cite{10.1007/978-3-031-81221-7_23}. It encodes semantic relationships between concepts extracted from economic, geopolitical, and policy documents. This KG is intentionally sparse, offering a clearer view of performance differences among baseline methods. Since this dataset is private and was not part of the training data for the LLMs used in its construction, we consider it private, ensuring that the models had no prior exposure to its content.
   % Socio-Economic: is a private LLM\fb{what does it mean?} generated dataset that represents semantic relationships between concepts from economic, geopolitical, or policy documents, this is a sparse dataset and its results should give a better insight of performance difference between our baselines since we can be sure that the LLMs we used did not have access to the dataset during their training.
\end{itemize}

%To reduce execution time, we evaluated our approach on sampled subsets of the knowledge graphs described above. Table~\ref{tab:dataset_stats} presents statistics for each dataset. Notably, we include the CoDEx-S and CoDEx-L datasets\cite{safavi2020codex} in the experiments although we mainly focused on CoDEx-M.

%To reduce the execution time, we tested our approach on samples of the knowledge graphs described above. Table \ref{tab:dataset_stats} presents statistics for each dataset. Notice that we have included in the table the statistics of the Covid-Fact dataset \cite{ezzabady2024towards} used in our motivating example. However, because of its limited size and sparsity, we did not include it in the main experimental evaluation.

%\textcolor{blue}{
%\mourad{I'm a bit confused here:
%(1) why a sample, what's the justification? What you said may be easily attacked by reviewers.
%(2) focus on CoDEx-S ... but experiments are done on all datasets!??
%(3) you need to better explain which dataset was used and where and why.}
Since LLM-based processing is time-consuming, we evaluate the LLM validation stage of our method on a sample of 500 generated candidates for each of the previously mentioned datasets. Table~\ref{tab:dataset_stats} reports the statistics of these datasets.
%We also include the CoDEx-S and CoDEx-L datasets~\cite{safavi2020codex}, although our main focus is on CoDEx-M. In addition, 
The table presents statistics for the Covid-Fact dataset~\cite{ezzabady2024towards} used in our motivating example. Due to its limited size, this dataset was not considered in the main experimental evaluation.
%}

\begin{table}[h!]
\centering
\begin{tabular}{lrrrr}
\hline
\textbf{Dataset} & \textbf{Relations} & \textbf{Entities} & \textbf{Triples}  \\
\hline
Socio-economic & 17,175 & 33,563 & 64,417 \\
CoDEx-M       & 49     & 16,759 & 60,000 \\
FB15K-237      & 29     & 12,993 & 59,270 \\
WN18RR & 11 & 40,943 & 93,003 \\
Covid-Fact & 28 & 1416 & 908 \\
\hline
\end{tabular}
\caption{Statistics of the datasets used in our experiments } 
\label{tab:dataset_stats}
\end{table}

%\subsection{Evaluation metrics}
To assess the performance of OMNIA, we adopt standard metrics commonly used in binary classification tasks, as our formulation of triple classification follows this setting. Specifically, we report the following metrics: accuracy, precision, recall, and F1-score. 
We define the standard terms as follows: 
(i) TP (True Positives) refer to correctly identified missing triples, 
(ii) TN (True Negatives) to correctly identified non-missing triples, 
(iii) FP (False Positives) to non-missing triples incorrectly classified as missing, and 
(vi) FN (False Negatives) to missing triples incorrectly classified as non missing.

\noindent \textbf{Ground Truth.} 
To compute the evaluation metrics, we simulate missing triples by removing  uniformly at random 20\% of the triples from each dataset, without conditioning on entities or relations. The remaining 80\% form the observed graph used for candidate generation, while the removed triples are treated as valid but unobserved facts to be recovered. 
This protocol serves as a controlled proxy for real-world incompleteness, where valid triples may be absent due to extraction failures or limited observations. 
To ensure robustness, we repeat the random removal five times and report averaged results.
Although random removal does not explicitly model all forms of implicit semantics found in LLM-generated KGs, it captures the key challenge of inferring true missing triples from existing structure and context. 
%For example, in CoDEx-M, a sampled set of 60k triples is split into 48k for generation and 12k for recovery.

%To be able to compute the above metrics, we apply a preprocessing step to simulate missing triples by randomly removing 20\% of the triples from each dataset. The remaining 80\% are used to generate candidate triples. For example, in the case of CoDEx-M, the sampled set of 60k triples is reduced to 48k for generation and 12k for recovery. \textcolor{red}{UPDATE [R2O3]}

\subsection{Baselines}
\label{subsection:baselineclass}
%To evaluate the performance of OMNIA, and demonstrate the effectiveness of each of its steps, we conducted experiments across the knowledge graphs described above. 
%A preprocessing script will drop 20\% of the dataset to simulate the missing triples to be recovered. we will only work with the remaining 80\% to generate our candidates ; in the case of codex-M, our sample that contains 60k triples will be reduced to 48k triples to work with and 12k triples to recover.
%First, we compare our generation method with an exhaustive search of all possible candidates in section 6.1. We then study the performance of our filtering components, its necessity and importance for OMNIA in section 6.2. Finally,  we compare the evaluation of the triple candidates (triple classification task) with three main classes of baselines, embedding based approaches, pretrained language model based approaches and LLM based approaches. For this experiment, we sampled 500 test triples, evenly distributed, for each dataset. We use the same test set for all the methods.
We compare OMNIA against three main classes of baseline methods:
\begin{itemize}
    \item \textbf{KGE-based:} We consider four representative KG embedding models: 
    TransE~\cite{DBLP:conf/nips/BordesUGWY13}, 
    DistMult~\cite{yang2015embeddingentitiesrelationslearning}, 
    ComplEx~\cite{trouillon2016complexembeddingssimplelink}, and 
    RotatE~\cite{sun2019rotateknowledgegraphembedding}. 
    For each model, we automatically select a classification threshold that maximizes the F1 score on a validation set.
    %Four knowledge graph embedding methods were tested on the same test set, these are TransE\cite{DBLP:conf/nips/BordesUGWY13}, Distmult\cite{yang2015embeddingentitiesrelationslearning}, ComplEx \cite{trouillon2016complexembeddingssimplelink}, RotatE\cite{sun2019rotateknowledgegraphembedding}. We automatically select a threshold that achieve the best F1 score on a defined validation set.
    
    \item \textbf{PLM-based:}  
    We include KG-BERT~\cite{yao2019kgbertbertknowledgegraph}, a widely used model for triple classification tasks that leverages 
    pre-trained language models. KG-BERT reformulates KG triples as natural language sequences and uses BERT to model their plausibility, enabling it to capture rich contextual and semantic information beyond traditional embedding-based methods. %as well as ExBert\cite{Jaradeh_2021} a model that reformulates knowledge graph triples into textual sequences and utilizes a pre-trained transformer language model to assess their plausibility.%We selected KG-Bert\cite{yao2019kgbertbertknowledgegraph} for our experiment since this model is a classic reference in related works and focuses on triple classification.
    
    \item \textbf{LLM-based:} For LLM based-method, we compare our method OMNIA to MuKDC~\cite{li2024llm} by evaluating MuKDC using the same evaluation protocol as OMNIA. We implemented MuKDC with a local Llava model using the Ollama library. 
    %These baselines correspond to the candidates' validation step of OMNIA, following the sentence-based and triple-based scenarios presented earlier. For each scenario, we evaluate the three prompting strategies: zero-shot, in-context and RAG. All experiments are conducted using the Mistral-7B model \cite{jiang2023mistral}. In the RAG setting, we compute embeddings using MPNet \cite{song2020mpnet}, a powerful sentence transformer, and FAISS \cite{johnson2019billion} to retrieve the most similar reference triples.

\begin{comment}
Existing LLM-based approaches for KG completion typically address tasks such as link prediction or KG construction. They are therefore not directly comparable to our task, which focuses on validating candidate triples through contextual evaluation. 
Consequently, we focus our baselines on OMNIA’s own variants, to allow a comparison across prompting strategies and input (triple-based and sentence-based)  representations.
\end{comment}
  %We separate LLM based KGC method into 3 subcategories zero-shot prompting, in-context and RAG-based. We used Mistral-7B\cite{jiang2023mistral} model as LLM to experiment with each scenario and case of OMNIA. We computed embeddings using our method with the MPNet \cite{song2020mpnet}, a powerful sentence transformer, and employed FAISS\cite{johnson2019billion} to retrieve the most similar triples.

\end{itemize}

\subsection{Implementation and Detail Settings}
\label{subsection:settings}
We use the Mistral-7B model \cite{jiang2023mistral} in all experiments. In the RAG setting, embeddings are computed with MPNet \cite{song2020mpnet}, and FAISS \cite{johnson2019billion} is used to retrieve the most similar reference triples.

The experimental evaluation was carried out on the Grid'5000 platform~\cite{balouek2013adding}, on the Graffiti node of the Nancy cluster. This node is equipped with an Intel Xeon Silver 4110 CPU, a GeForce RTX 2080 Ti GPU, and 128 GB of RAM.
%All experiments were conducted on the Grid'5000~\cite{balouek2013adding}, specifically on the Graffiti node of the Nancy cluster, equipped with an Intel Xeon Silver 4110 CPU, a GeForce RTX 2080 Ti GPU, and 128 GB of RAM. 
An exception was made for the exhaustive search of candidate triples, which required more memory and was therefore executed on the Grosminet node within the same cluster. Grosminet features an Intel Xeon Gold 6240L CPU, 6144 GB of RAM, and no GPU.

In the validation step of OMNIA, we first filter the generated candidate triples before LLM-based evaluation. For this, we used PyKEE\cite{ali2021pykeen} to implement TransE \cite{bordes2013translating}. To select the optimal filtering threshold, we evaluated four possible values on a validation set: median+0.5 , mean +0.5,  median - 0.5 and mean -0.5. The threshold yielding the highest F1-score was retained for use during evaluation.
In the RAG case of the LLM-based validation, we used the LangChain\cite{langchain} framework to manage both retrieval and language generation. To ensure semantically meaningful chunks, we employed the SemanticChunker alongside HuggingFaceEmbeddings\cite{song2020mpnet}. The resulting embeddings were indexed using FAISS\cite{johnson2019billion}, enabling efficient similarity-based retrieval during the generation process.

OMNIA uses Mistral-7B by default alongside LangChain, for all LLM-based prompting cases. This model offers a strong balance between performance, efficiency, and scalability, making it well-suited for our validation scenarios.

For the KG embedding (KGE) baselines, we rely on PyKEEN to implement TransE, DistMult, ComplEx, and RotatE. For KG-BERT, we used the official implementation available on GitHub\footnote{\url{https://github.com/yao8839836/kg-bert}}. All baselines were evaluated on the same test set, which includes 500 triples per dataset, evenly distributed between true and false missing triples. In addition, the KGE and PLM-based methods were provided with a validation set consisting of 250 triples, evenly split between positive and negative examples, for selecting the optimal classification threshold. All models were trained on the same set of candidate triples generated by OMNIA, ensuring a consistent and fair comparison across methods.

We now present the results of our experiments in the next subsections. All experiments are repeated five times using different random splits, and the reported results are averaged across runs.
%Section~\ref{subsection:generation} compares OMNIA’s candidate generation step to an exhaustive search of all possible candidates. Section~\ref{subsection:filtering} evaluates the effectiveness and impact of the filtering sub-step within OMNIA’s validation stage. Section\ref{subsection:baselines}  presents our evaluation of the LLM-based validation using the sentence-based and triple-based scenarios compared to the three baseline families: KGE, PLM, and LLM. Section~\ref{subsection:prompting} shows the performance of the different prompting strategies. Finally, Section~\ref{subsection:hyper} shows our analysis of the hyperparameters used by OMNIA.

\subsection{Results}
\label{subsection:res}
%\textcolor{blue}{This section presents the results  across OMNIA's main components, namely triple generation, candidate filtering, and LLM-based validation. }
\subsubsection{OMNIA's Triple Generation}
\label{subsection:generation}

\begin{table*}[t]
\centering
\setlength{\tabcolsep}{4pt}
\begin{tabular}{|l|l|l|l|l|l|l|l|l|l|l|}
\hline
\textbf{dataset} & \textbf{UE\_O} & \textbf{UE\_G} & \textbf{UR\_O} & \textbf{UR\_G} & \textbf{nb\_candidates} & \textbf{nb\_missings} & \textbf{TP} & \textbf{nb\_miss\_brut} & \textbf{nb\_gen\_brut} \\ \hline
CoDEx-M & 16759 & 16431 & 49 & 48 & 9047869 & 11996 & 70.65\% & 95.98\% & 12958932528 \\ \hline
Socio-economic & 33563 & 29860 & 17175 & 14710 & 6494589 & 12347 & 9.6\% & 55.76\% & 13115724316000 \\ \hline
FB15K237 & 12993 & 12610 & 29 & 29 & 4885007 & 11854 & 49.08\% & 96.12\% & 4611350900\\ \hline
\end{tabular}
\caption{Comparison of dataset statistics and generation performance. U: unique , E: entities , R: relations , O: original dataset size, G: Left for generation , TP\_OMNIA: \% of missing triples successfully generated by OMNIA, TP\_brut:\% of missing triples successfully generated by an exhaustive search, gen\_brut: Total number of candidate triples generated by exhaustive search.}
\label{tab:dataset_stats_2}
\end{table*}

In this section, we answer (Q1) related to the efficiency and effectiveness of OMNIA's candidate generation to identify relevant missing triples. We evaluate our proposed approach prior to any downstream filtering or validation. To provide a meaningful comparison, we consider an exhaustive triple generation strategy. Although we did not execute the exhaustive approach in full for all the datasets due to computational constraints, we estimate, on a per-dataset basis, the number of missing triples that would not be generated as a result of discarding certain entities and relations during the experiment setup.

%(answer for Q1) In this section, we evaluate the effectiveness of our proposed approach for generating missing triples, prior to any downstream filtering or evaluation. We also compare our method to an exhaustive triple generation strategy.  While we did not execute the exhaustive approach in full due to computational constraints, we estimated, for each dataset, the number of triples that would have been missed as a result of discarding certain entities and relations during the experiment setup.

The comparative results are presented in Table~\ref{tab:dataset_stats_2}. Although our approach generates fewer candidate triples than an exhaustive generation strategy, it offers a favorable trade-off between scalability and coverage. OMNIA prioritizes computational efficiency and scalability, focusing on high-quality candidates rather than brute-force enumeration. To underscore the impracticality of exhaustive generation, we executed it on a sample of the CoDEx-M dataset with the workload distributed across 10 processes. This run required nearly two hours and consumed approximately 2.6 terabytes of RAM. In contrast, OMNIA completes the same task in just 10 minutes on a single process, well within the limits of 8 gigabytes of RAM.
%trade-off leveraging scalability over exhaustive generation. OMNIA prioritizes the computational efficiency and scalability, focusing on high-quality candidates rather than brute-force enumeration. 
%Although our method generates fewer candidate triples than exhaustive generation, it is important to emphasize that this is a deliberate design choice. OMNIA prioritizes computational efficiency and scalability.
%To underscore the impracticality of exhaustive generation, we executed it on a sample of the CodEx-M dataset with the workload distributed across 10 processes. This run required nearly two hours and consumed approximately 2.6 terabytes of RAM. In contrast, OMNIA completes the same task in just 10 minutes on a single process well within the limits of 8 gigabytes of RAM.

For instance and as presented in Table~\ref{tab:dataset_stats_2}, in the CodEx-M sample, the number of unique entities decreased from 16,759 (UE\_O) to 16,431 (UE\_G), and relations from 49 (UR\_O) to 48 (UR\_G) after dropping a set of triples used as missing triples for evaluation. As a result, an exhaustive search would generate $16431 \times 16431 \times 48 = 12958932528$ triples, among which 95.98\% of the missing triples are generated (TP\_brut). In comparison, OMNIA generates 9,047,869 candidate triples; approximately 1,400 times fewer than the exhaustive search, while still generating 70.65\% of the missing triples  (TP\_OMNIA). This experiment demonstrates OMNIA’s ability to drastically reduce the candidate space while still successfully generating a substantial portion of the known missing triples.

To better understand the generation performance on sparse versus dense knowledge graphs, we extended the experiment on the socio-economic dataset. Specifically, we created a denser subset, referred to as \textit{Socio2}, containing 12,000 triples in which each entity and relation appears at least three times. OMNIA’s generation performance improved from 9.6\% on the original sparse dataset to 30\% on Socio2. We further evaluated OMNIA on the full CodEx-M and CodEx-S datasets, where it successfully generated over 97\% and 98.3\% of the missing triples, respectively. These results confirm that OMNIA is particularly effective in dense KGs, where entities and relations occur with greater frequency.

The Socio-economics dataset is a sparse KG where structural patterns are less frequent. Thus, the clustering-based candidate generation produces fewer valid candidates, which may result in lower recall before the validation stage. This behavior mainly concerns the candidate generation step, while the LLM-based validation remains effective once candidates are available. Potential mitigation strategies for sparse graphs are discussed as future work in Sectio~\ref{sec:conclusion}.

\subsubsection{OMNIA’s Candidate Filtering}
\label{subsection:filtering}

As presented previously, filtering plays a crucial role by reducing the number of candidate triples presented to the LLM-based evaluation. It aims to retain only the most promising triples from the large set generated during the first stage, thereby improving computational efficiency without sacrificing too much recall. Table~\ref{filtering_table} shows the performance of this filtering step across different datasets. In the table, we  report the reduction ratio, the total number of missing triples, the number of true candidates before filtering (TC), and the number of true candidates after filtering (TCF).
For the CodEx-M sample, OMNIA initially generates 9 Million candidates that contain 8,476 triples that are actual missing triples (TC). 
After filtering, the candidates are reduced by 71.08\% while introducing a loss in TC also reducing it to 5,024 triples (TCF). 
The total number of missing triples in this case is 11,997. On the FB15K237 dataset, we observe a 41.76\% reduction in candidates, also introducing a loss, from 5,818 (TC) to 3,836 (TCF), in a total of 11,854 missing triples. 
The same trend is seen on the sparse Socio-economic dataset, where the number of candidates was reduced by 70\% for a 34\% loss of TC.
While the filtering step does result in the loss of some valid triples, the overall computational gain is substantial. Despite a filtering accuracy of around 60\%, this step remains beneficial, as it significantly reduces the load for LLM-based evaluation. Further improvements, such as better embedding techniques or refined training, are left as future work.
%. For the CoDEx-M sample , after generating 9M candidates, the filtering module reduced this amount into a 2.5M candidates, decreasing the TC from 8476 out of 11996 to 5024 TCF. a 71\% reduction in candidates for a 40\% loss in TC. The results are quite similar for all the datasets, on FB15K237 we notice a 41.76\% reduction in candidates for 34\% loss of TC.

\begin{table}[h]
\centering
\begin{tabular}{|c|c|c|c|c|}
\hline
\textbf{Dataset} & \textbf{Red. Ratio} & \textbf{Missing} & \textbf{TC} & \textbf{TCF} \\ \hline
CoDEx-M & 71.08\% & 11997 & 8476 & 5024 \\ \hline
Socio-economic & 70\% & 12347 & 1186 & 607 \\ \hline
FB15K237 & 41.76\% & 11854 & 5818 & 3836 \\ \hline
\end{tabular}
\caption{Filtering performance over different datasets}
\label{filtering_table}
\end{table}
\setlength{\tabcolsep}{4pt}

\subsubsection{Effectiveness of OMNIA compared to baselines} 
\label{subsection:baselines}

%\textcolor{red}{complete the results of plm based in the table}
In this section, we answer question Q2 related to the effectiveness of OMNIA’s evaluation model compared to existing baseline methods. 
We evaluate the performance of OMNIA's LLM-based evaluation on the task of triple classification using standard metrics: accuracy, precision, recall and F1-score. We present the F1-score in Table~\ref{tab:f1_results}, as it provides a comprehensive summary of the model’s performance across all metrics.
As discussed earlier, we evaluate OMNIA against representative baselines, including knowledge graph embedding (KGE) models (TransE, DistMult, ComplEx, RotatE), the PLM-based KG-BERT, and the recent LLM-based method MuKDC. OMNIA’s evaluation component is assessed under both sentence-based and triple-based prompting strategies, using three prompting cases: zero-shot, in-context and retrieval-augmented generation (RAG).

Overall, OMNIA demonstrates strong performance across all datasets. In particular, the RAG-based variant of OMNIA consistently achieves the highest F1-scores on FB15k-237 (0.86), CoDEx-M (0.91), and WN18RR (0.87), outperforming all baseline models. These gains highlight the benefit of retrieving relevant context to support triple validation by the LLM. For example, OMNIA outperforms TransE on FB15k-237 (F1: 0.86 vs. 0.74) and CoDEx-M (F1: 0.91 vs. 0.68), and outperforms DistMult on WN18RR (F1: 0.87 vs. 0.77).
On the Socio-Economic dataset,  OMNIA performs at a comparable level to the baselines, with an F1-score of 0.68, while the best KGE baseline, DistMult, achieves 0.74 and KG-Bert achieves 0.73.

The LLM-based baseline MuKDC has low overall F1-scores across all datasets,  with high recall and low precision. This behavior can be attributed to its aggressive candidate generation strategy, which prioritizes coverage over accuracy, leading to a very high recall but poor precision. In contrast, OMNIA maintains a better balance between recall and precision, resulting in substantially higher overall performance.

%\textcolor{red}{Although MuKDC appears to have low overall performance, it achieves high recall with low precision. This is explained by its aggressive candidate generation strategy, which favors coverage over accuracy which result in very high recall but rather poor precision.}

These results highlight the strength of our evaluation strategy and confirm the added value of using LLMs with context retrieval for the triple classification task. Since the removal is uniformly random, the removed triples include both cases that can be recovered from structural cues and cases that cannot. Using standard benchmarks also shows that our method generalizes beyond LLM-generated KGs. Moreover, OMNIA outperforms state-of-the-art structural methods such as TransE and DistMult, which rely on graph structure but not semantics. This shows that our approach is better at capturing semantic signals beyond pure structural patterns.
%\textcolor{red}{These results highlight the strength of our evaluation strategy and confirm the added value of using LLMs with context retrieval for the triple classification task. Since the removal is uniformly random, the removed triples include both cases that can be recovered from structural cues and those that cannot. We repeat the random removal five times to ensure robustness. Using standard benchmarks also shows that our method generalizes beyond LLM-generated KGs. Moreover, OMNIA outperforms state-of-the-art structural methods such as TransE and DistMult, which rely on graph structure but not semantics. This shows that our approach better captures semantic signals beyond pure structural patterns.}
%These results highlight the strength of our evaluation strategy and confirm the added value of using LLMs with context retrieval for the triple classification task.
\textcolor{red}{}
%(Answer for Q2) In this section we compute the accuracy, precision, recall and F1-score of our evaluation component to fully capture the ability of OMNIA to classify the generated candidate triples. The results of this experiment are reported in table\ref{tab: main_results}.
%Overall OMNIA outperforms all the baselines on all the datasets as it hits an f1-score of 0.86, 0.91 and 0.87 on FB15k-237, CoDEx-M and WN18RR respectively. while the best f1 score in our baselines was TransE's 0.74, TransE's 0.68 and DistMilt's 0.77 on the same datasets.\textcolor{blue}{ This is due to the use of RAG for finding the context as it provided a context that suits best the candidate and is usually enough for the LLM to make its decision.} OMNIA does perform at the same level as the baselines on socio-Economic dataset, as it scores 0.68 while the best f1-score in our baselines is DistMult's 0.74. The results prove the superiority of our approach using RAG at triple classification.

{\small
\renewcommand{\arraystretch}{1.2}
\setlength{\tabcolsep}{1.3pt}
\begin{table}[t]
\centering
\begin{tabular}{llcccc}
\toprule
\textbf{Category} & \textbf{Model} & \textbf{FB15k-237} & \textbf{CoDEx-M} & \textbf{WN18RR} & \textbf{Socio-Economic} \\
\midrule
\textbf{KGE} 
& TransE & 0.74 & 0.682 & 0.68 & 0.72 \\
& DistMult & 0.6709 & 0.674 & 0.77 & \textbf{0.737} \\
& ComplEx & 0.6661 & 0.665 & 0.67 & 0.66 \\
& RotatE & 0.7285 & 0.662 & 0.75 & 0.717 \\
\midrule
\textbf{PLM-based} 
& KG-Bert & 0.57 & 0.542 & 0.62 & \underline{0.727} \\
\midrule
\textbf{LLM-based}
& MuKDC & 0.192 & 0.273 & 0.229 & 0.224 \\
\midrule
\multirow{3}{*}{\shortstack{\textbf{OMNIA}\\\textbf{Sentences}}}
& Zero-shot & 0.68 & 0.68 & 0.63 & 0.59 \\
& In-context & 0.65 & 0.69 & 0.66 & 0.66 \\
& RAG & \underline{0.85} & \textbf{0.91} & 0.72 & 0.678 \\
\midrule
\multirow{3}{*}{\shortstack{\textbf{OMNIA}\\\textbf{Triples}}}
& Zero-shot & 0.75 & 0.62 & 0.74 & 0.579 \\
& In-context & 0.61 & 0.65 & \underline{0.83} & 0.649 \\
& RAG & \textbf{0.86} & \underline{0.84} & \textbf{0.87} & 0.584 \\
\bottomrule
\end{tabular}
\caption{Performance (F1-score) comparison of OMNIA against baselines across four datasets.  \textbf{Bold} indicates the best per dataset; \underline{underlined} indicates the second-best. OMNIA (with its different variants) performs the best across all datasets with a big margin (12 percentage points for FB15k-237, 23.6 for CoDEx-M, and 10 for WN18RR) except "Socio-Economic", where it ranks third.}
\label{tab:f1_results}
\end{table}
}

\subsection{Effect of Sentence-based vs. Triple-based Prompting}
\label{subsection:prompting}

%\subsection{Triple or sentence representation for the candidate}

In this section, we answer question Q3 that investigates how the the representation of the candidate impact its evaluation by the LLM. 
Specifically, we compare the two scenarios in OMNIA’s LLM-based evaluation, (i) candidate triples expressed in natural language sentences (OMNIA Sentences), 
and (ii) structured triples (OMNIA Triples).
LLMs are primarily trained to process natural language sentences rather than structured triples. Consequently, representing a candidate triple as a sentence is expected to result in better performance.%LLMs are trained to work on full sentences, not structured triples. Therefore, we expect better performance when a triple is expressed as a sentence. This observation directly motivates our design choice to include a sentence-based validation option alongside triple-based validation in OMNIA.

The results, presented in Table \ref{tab:f1_results}, show that OMNIA sentences variant outperforms OMNIA triples variant across CoDEx-M and Socio-Economic, regardless of the prompting strategy (zero-shot, in-context, or RAG), confirming the benefit of sentence-based validation in these settings. 
For example, on CodEx-M, both accuracy and F1-score improved by over 5\% using a sentence representation.
However, on WN18RR, OMNIA triples yield better results. This can be attributed to the dataset’s unique entity and relation representations (e.g., “X.nn.02”), which hinder the LLM’s ability to correctly interpret triples transformed into sentences. For instance, in the WN18RR dataset, the  incorrect candidate triple (\textit{theophrastaceae.n.01 , \_member\_meronym, pistacia.n.01}) was classified as False by the LLM (Mistral\_7B \cite{jiang2023mistral}) when given the triple. However, when it was reformulated into the natural language sentence "\textit{A member of the Theophrastaceae is a pistachio tree}", the model revised its judgment and classified it as True. This underscores the importance of supporting both representations to enhance the generalizability of the solution across datasets.
However, on WN18RR, OMNIA triples yield better results. This can be attributed to the dataset’s unique entity and relation representations (e.g., “X.nn.02”), which hinder the LLM’s ability to correctly interpret triples transformed into sentences. For instance, in the WN18RR dataset, the  non-valid candidate triple (\textit{theophrastaceae.n.01 , \_member\_meronym, pistacia.n.01}) was classified as False by the LLM (Mistral\_7B \cite{jiang2023mistral}) when given the triple. However, when it was reformulated into the natural language sentence "\textit{A member of the Theophrastaceae is a pistachio tree}", the model revised its judgment and classified it as True. This underscores the importance of supporting both representations to enhance the solution's generalizability across datasets.

%In this section, we compare the results of OMNIA when given a candidate to evaluate shaped as a sentence(OMNIA sentences) with OMNIA when given a candidate shaped as a triple(OMNIA triples). we report our results in table \ref{tab: main_results}. We notice that Omnia sentences performs better on codex-M, FB15K237 and socio-economic regardless of the level of contextual information provided for example on codex-M dataset both accuracy and F1 score increased by 5\%. However, OMNIA triples does perform better on WN18RR dataset, and the reason for that is uniqueness -compared to the other datasets- of the representation of its entities and relations. An entity "X.nn.02" in WN18RR refers to the second sense of the word X. Such representation confuses the LLM while transforming it into a sentence. This experiment is important to test and identify the strength of our method as well as validate our choice of having both representations in OMNIA in order to make it adaptable to other use cases than LLM-generated KGs.

\noindent \textbf{Impact of Prompt Templates.}
%\textcolor{red}{i rewrite this subsection, but something is missing: what are the prompt templates?is it the case for all datasets? specify the dataset for which we are using the 500 triples... can we add a table shwing the percentage of the transformed sentences for each dataset?}
We also experimented with several prompt templates to transform triples into sentences.
We found that LLMs are prone to rewriting semantically non valid triples into seemingly valid sentences if not explicitly instructed otherwise.
This behavior is problematic, as the evaluation relies on preserving the intended meaning of the original triples.
%When explicitly instructed not to correct the input triples, all 500 triples in our test sample were correctly transformed into sentences that preserved their original semantics. However, without such explicit instructions, approximately 8\% of the transformed sentences conveyed the opposite meaning of the original triples.
To illustrate this issue, we compare a simple prompt that lacks detailed instructions with an explicit prompt that clearly guides the model to preserve the original semantics of the triple. The simple prompt used is:
%\textcolor{orange}{The following prompt would be the simple prompt when we do not explicit what to not do.}
\begin{tcolorbox}[
  colback=white,
  colframe=black!75,
  boxrule=0.5pt,
  sharp corners=south,
  rounded corners=north,
  left=4pt,
  right=4pt,
  top=4pt,
  bottom=4pt,
   before skip=4pt,
  after skip=4pt,
  enhanced,fontupper=\small
]
\textbf{Simple prompt:} \\
Transform the following triple into a sentence: \{head\} \{relation\} \{tail\} \\
Answer: Give the sentence.
\end{tcolorbox}

This prompt provides no constraint on whether the model should maintain the validity of the triple, as explained above. To address this issue, we design the following explicit prompt, which gives clear, step-by-step instructions to avoid unintended rewriting.
%\textcolor{orange}{The following prompt would be the explicit prompt when we explicit what to not do.}
\begin{tcolorbox}[
  colback=white,
  colframe=black!75,
  boxrule=0.5pt,
  sharp corners=south,
  rounded corners=north,
  left=4pt,
  right=4pt,
  top=4pt,
  bottom=4pt,
  before skip=4pt,
  after skip=4pt,
  enhanced,fontupper=\small]
\textbf{Explicit prompt:}\\
Your task is to only transform the triple into a sentence, no matter if it is valid or not.\\
A triple has a head, a relation, and a tail.\\
Transform the following triple: \texttt{\{head\} \{relation\} \{tail\}} as is.\\
Do not rephrase non-valid facts or use negative constructions.\\
Answer: Give the sentence.
\end{tcolorbox}
\begin{comment}
    
\begin{tcolorbox}[colback=white,colframe=black!75,boxrule=0.5pt,sharp corners]
\textbf{Explicit prompt:}
\begin{enumerate}
    \item Your job is only to translate triple into sentence, no matter if it is correct or not
    \item A triple is two entities (head and tail) linked by a relation
    \item Transform the following triples into a sentence: '{head} {relation} {tail}'
    \item If the triple present incorrect fact, still translate this as it is
    \item Do not make negative sentence 
\end{enumerate}
Answer: Give the sentence.
\end{tcolorbox}
\end{comment}
We evaluate these two prompts on a sample of 500 candidate triples generated from the CoDEx-M dataset. The results in Table~\ref{tab:prompt_sent} show that without explicit instructions, the LLM incorrectly rewrites 8\% of the triples, altering their original meaning. With an explicit prompt, it consistently preserves semantics, achieving 100\% correct transformations in this dataset.
%The results of this evaluation on a sample of 500 candidates triples generated from CoDEx-M dataset is shown in table \ref{tab:prompt_sent}.
\begin{table}[h]
    \centering
    \begin{tabular}{|c|c|c|}
    \hline
         \textbf{Method} & \textbf{Valid trans.} & \textbf{Valid trans. ratio}\\
         \hline
         Simple prompt & 462 & 0.92\\
         Explicit prompt & 500 & 1\\
        \hline
    \end{tabular}
    \caption{Results of simple and explicit prompts for sentence transformation on 500 candidates triples from CoDEx-M dataset}
    \label{tab:prompt_sent}
\end{table}
%We have found that LLM are prone to transform semantically incorrect triple into semantically correct sentences if not explicitly instructed not to.
%It is an issue, as the LLM evaluate When explicitly instructed to not correct incorrect triple into correct sentence, on a sample of 500 triples, all of triple were correctly transformed into sentences.
%When we do not give theses explicits instructions, on the same sample of 500 triples, around 8\% are transformed into the opposite of what the triple meant.

\begin{figure*}[th]
    \centering
    \begin{subfigure}{0.24\textwidth}
        \centering
        \includegraphics[width=\linewidth]{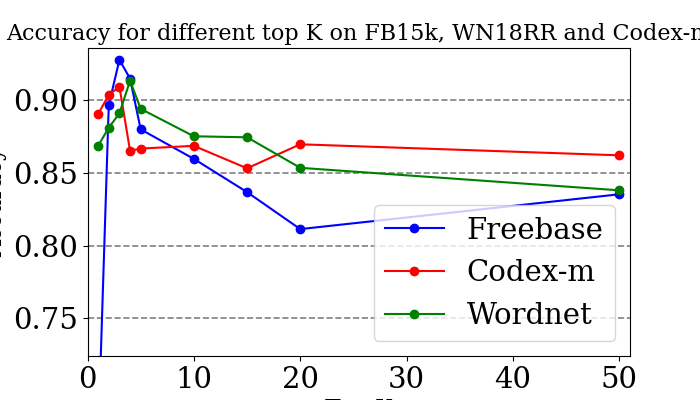}
        \caption{Accuracy}
        \label{fig:top_k_accuracy}
    \end{subfigure}
    \hfill
    \begin{subfigure}{0.24\textwidth}
        \centering
        \includegraphics[width=\linewidth]{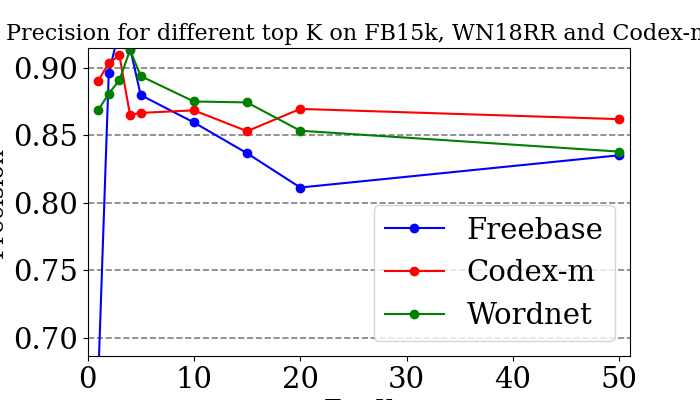}
        \caption{Precision}
        \label{fig:top_k_precision}
    \end{subfigure}
    \hfill
    \begin{subfigure}{0.24\textwidth}
        \centering
        \includegraphics[width=\linewidth]{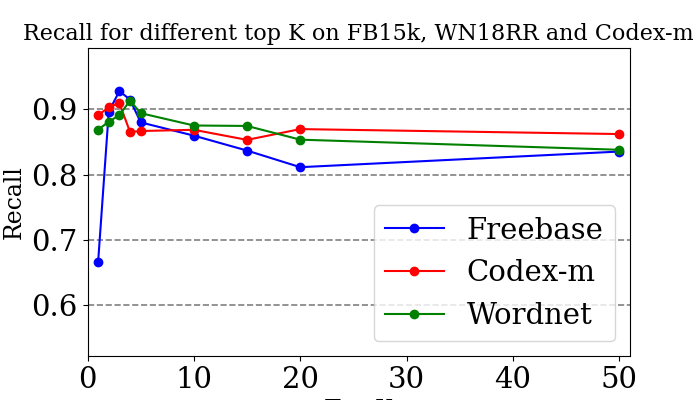}
        \caption{Recall}
        \label{fig:top_k_recall}
    \end{subfigure}
    \hfill
    \begin{subfigure}{0.24\textwidth}
        \centering
        \includegraphics[width=\linewidth]{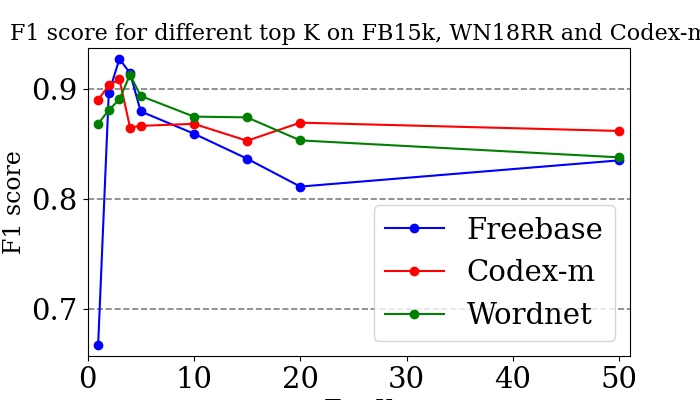}
        \caption{F1 score}
        \label{fig:top_k_F1}
    \end{subfigure}
    \hfill 
    \caption{Scores for different top-K on FB15K-237, WN18RR, CoDEx-M and Socio-economic}
    \label{fig:score_top_k}
\end{figure*}

\subsection{Hyperparameter Analysis}
\label{subsection:hyper}
%\fb{this is not interesting, a qualitative analysis instead would be better}

In this section, we answer question Q4, by analyzing the effect of the hyperparameter top-$k$ in OMNIA’s prompting strategies, particularly in the RAG setting. 
In a RAG framework, the top-$k$ determines the number of similar triples retrieved from the KG that will be then added as a context for the prompt presented to the  LLM.
%-based validation. 
We evaluate how varying this parameter affects performance on the triple-based scenario.

We experimented with top-k values in $\{1, 2, 3, 4, 5, 10, 15, 20, 50\}$ and computed our metrics, namely accuracy, precision, recall, and F1-score for datasets WN18RR, FB15k-237 and CoDEx-M. 
For example, on CoDEx-M, we observed that setting top-$k$ = 50 introduces many irrelevant triples, making it harder for the model to select the correct answer. In contrast, top-$k$ = 2 yields mostly relevant triples, improving accuracy. 
%\textcolor{orange}{For example, on CoDEx-M for the triple candidate (Florian David Fitz, occupation, restaurateur) setting top-$k$ = 50 introduces many irrelevant triples, the model would see it as a correct triple. In contrast, for top-$k$ = 2, as it only see few triples, it correctly assess it as incorrect.}
As shown in Figure \ref{fig:score_top_k}, the F1-score peaks at top-$k$ = 3 across all datasets. It drops significantly for top-$k$ = 1 compared to top-$k$ = 2 and top-$k$ = 3. Moreover, for values above 5, F1-scores dropped substantially due to the inclusion of less relevant context. We conclude from this experiment that the optimal range for top-$k$ lies between 2 and 4.  Based on these results, we selected top-$k$ = 2 as the optimal setting for our experiments.
%(answer for Q4) In this section, we evaluate the significance of the hyper-parameter top-$k$ in our queries for our RAG-based method in triple scenario.
%On RAG, the top-$k$ value is the number of most similar triples that we choose to give as a context to our RAG system.

%We evaluated our method using accuracy, precision, recall, F1 score for top-$k$ values $\in \{1, 2, 3, 4, 5, 10, 15, 20, 50\}$.
%We conducted experiments on several datasets: WN18RR, FB15k-237 and CoDEx-M.

%\textcolor{orange}{For example, for CoDEx-M, for the triple (,,), we have found that for top-k = 50, we have a lot of irrelevant triple, hindering our model abilities to make the right choice; however for top-k = 2, we only got relevant triples, hence our method is getting the correct result.}

%The Accuracy, Precision, Recall and F1 Score by top-$k$ are shown in figure \ref{fig:score_top_k}.
%For every datasets, we observed that the F1 score reaches its highest value when top-$k=3$. Additionally, we found that for top-$k=1$, the F1 scores were significantly lower compared to top-$k=2$ and top-$k=3$. 
%Furthermore, for top-$k=5$ and higher values, the F1 scores dropped substantially.
%\textcolor{orange}{We observe that giving too many triples as context to the LLM would mess up the LLM reasoning }
%We can conclude from that experiment to the best value for top-$k$ is between 2 and 4, hence we choose to conduct our experiments using top-$k$ = 2.
\subsection{Generalizability}
We evaluate OMNIA’s performance across diverse Knowledge Graphs and types of missing triples to assess its generalizability beyond implicit semantic incompleteness.
Our evaluation includes lexical KGs such as WN18RR, which capture structured linguistic hierarchies; encyclopedic KGs such as CoDEx-M and FB15K-237, which contain dense and heterogeneous relational patterns; and a socio-economic KG, which is sparse and semantically rich, derived from unstructured textual data.
Although OMNIA was originally designed to recover missing triples in LLM-generated Knowledge Graphs, its architecture, which combines clustering, embedding-based filtering, and LLM-based validation, also allows it to handle incompleteness in traditional KGs.
These results show that OMNIA generalizes well beyond its initial scope and adapts effectively to heterogeneous KG completion scenarios.

\subsection{Efficiency}
\label{subsection:efficiency}
We evaluated the efficiency and scalability of OMNIA by measuring the average execution time of its main components on CoDEx-M and FB15k-237 samples of increasing sizes.
%\textcolor{red}{For this section, all experiments were run locally on a single machine on a CPU, without parallel or distributed execution.} 
Table~\ref{tab:time_generation_size} reports the results for the candidate generation and embedding-based filtering steps. 
Candidate generation is the most time-consuming operation, with execution time increasing faster than linearly as dataset size grows, consistent with the complexity analysis presented above.
However, since this step is performed offline, the cost remains acceptable.
In contrast, the embedding filtering step exhibits a near-linear trend with a small and stable overhead. 
These results demonstrate that OMNIA maintains practical execution times and scales efficiently even for larger samples, confirming its suitability for medium and large-scale Knowledge Graphs.

Table~\ref{tab:time_LLM} reports the average execution time of the LLM-based evaluation across different sample sizes of candidates. As expected, the execution time increases with the number of evaluated triples, reflecting the near-linear dependence of LLM inference on input size.
Although the execution time appears long, since each evaluation is independent of the others, it can be significantly reduced by distributing the workload across multiple machines and nodes, as parallel processing enables efficient scaling.

Among all prompting strategies, zero-shot evaluation is the fastest for both triple-based and sentence-based inputs, because it does not require any context retrieval. In-context prompting introduces moderate overhead due to the inclusion of a small set of examples, while RAG incurs the highest computational cost because of its additional retrieval stage.
For smaller sample sizes, the in-context strategy is notably more efficient than RAG. However, as the number of evaluated triples increases, their execution times converge, indicating that RAG’s overhead remains constant and that its runtime scales almost linearly with dataset size.
Overall, these results confirm that all prompting strategies exhibit consistent and scalable runtime behavior, runtime behavior, with RAG offering the best trade-off between contextual accuracy and computational efficiency.

\begin{table}[h!]
\centering
\caption{Average execution time (in seconds) of each step across different sample sizes (in thousands).}
\label{tab:time_generation_size}
\begin{tabular}{lccccc}
\toprule
\textbf{OMNIA's step} & \multicolumn{5}{c}{\textbf{Sample Size (in thousands)}} \\
\cmidrule(lr){2-6}
 & \textbf{10} & \textbf{20} & \textbf{30} & \textbf{40} & \textbf{50} \\
\midrule
Candidate generation& 6.08 & 26.47 & 72.42 & 160.24 & 304.83 \\
Embedding filtering& 13.59 & 15.16 & 19.47 & 24.30  & 30.53 \\
\bottomrule
\end{tabular}
\end{table}
\begin{comment}
\begin{figure}
    \centering
    \includegraphics[width=0.9\linewidth]{results/scalability-cand.png}
    \caption{Candidates generation and embedding filtering scalability}
    \label{fig:scalability-cand}
\end{figure}
\end{comment}

\begin{table}[h!]
\centering
\caption{Average execution time (in seconds) of LLM-based evaluation for different sample sizes.}
\label{tab:time_LLM}
\begin{tabular}{lcccc}
\toprule
\textbf{LLM Scenario} & \multicolumn{4}{c}{\textbf{Sample Size}} \\
\cmidrule(lr){2-5}
 & \textbf{50} & \textbf{100} & \textbf{250} & \textbf{500} \\
\midrule
\multicolumn{5}{c}{\textit{Triple-based Evaluation}} \\
\midrule
Zero-shot & 85.18 & 181.87 & 532.52 & 1014.35 \\
In-context & 138.59 & 264.61 & 672.2 & 1481.35 \\
RAG & 567.76 & 687.24 & 987.33 & 1489.83 \\
\midrule
\multicolumn{5}{c}{\textit{Sentence-based Evaluation}} \\
\midrule
Zero-shot & 133.91 & 326.7 & 756.07 & 1517.59 \\
In-context & 300.53 & 687.56 & 1712.83 & 3365.93 \\
RAG & 743.64 & 1057.81 & 1852.87 & 3156.91 \\
\bottomrule
\end{tabular}
\end{table}

We further compared OMNIA with standard embedding approaches using a sample of 50,000 triples from the CoDEx-M and FB15K-237 datasets.
This size was selected for consistency with our scalability analysis, ensuring a representative yet computationally feasible comparison.
Traditional embedding models such as TransE, RotatE, ComplEx, and DistMult completed training and scoring in 35 seconds to 53 seconds on average. %whereas OMNIA’s full pipeline required around 420,000 s on the same sample roughly 11,500 times slower. 
Although OMNIA is slower,  it provides significant performance improvement. It increases the F1-score by 23\% on CoDEx-M (from 0.68 to 0.91) and by 12\% on FB15K-237 (from 0.74 to 0.86) compared to the best embedding-based baseline.
These results confirm that OMNIA’s higher computational cost directly translates into better semantic accuracy and higher-quality triple validation, making it particularly effective in precision-critical KG completion scenarios.
\section{Conclusion}
\label{sec:conclusion}

In this paper, we introduce OMNIA, a two-stage approach for improving the completeness of knowledge graphs. OMNIA is broadly applicable to various types of KGs, while being particularly well suited to LLM-generated KGs, where some triples are semantically entailed by the source text but not explicitly extracted.%while also being particularly suited to address a specific class of missing triples found in LLM-generated KGs, those that are semantically entailed by the source text but not explicitly extracted. 

%OMNIA is based on an assumption drawn from a real-world use case and supported by manual quality assessment which revealed that many missing triples arise because entities sharing the same relation with a common entity are often similar and tend to share additional attributes. 
OMNIA first clusters entities that appear in similar relational contexts to generate candidate triples. These candidates are then filtered using a lightweight embedding-based model and validated by an LLM through different prompting strategies. We conducted extensive experiments across multiple datasets, demonstrating that OMNIA consistently improves triple classification performance over existing baselines, with gains of up to 23 percentage points in F1-score on several benchmarks.

%OMNIA's core idea stems for a real-world use case, validated by manual quality assessment. This assessment revealed that many missing triples occur because entities sharing a relation with a common entity often exhibit similarities and tend to share additional attributes.

%Building on the above insight, OMNIA first clusters such entities to generate new candidate triples. These candidates are then filtered using a lightweight embedding-based model and subsequently validated by a large language model using different prompting strategies. We conducted extensive experiments across a variety of datasets, showing that OMNIA significantly improves triple classification performance compared to existing baselines, achieving up to 23 percentage points gain in F1-score on several benchmarks. 

Our results  show that OMNIA is particularly effective on denser KGs, where structural patterns occur more frequently. On very sparse graphs, candidate generation becomes more challenging, as fewer valid candidates can be produced prior to validation, while the LLM-based validation remains effective once candidates are available.
To address this limitation, we plan to improve candidate generation for sparse datasets. Possible directions include automatically detecting sparse regions of a KG and applying alternative strategies, such as semantic similarity or LLM-assisted candidate generation when structural signals are insufficient. These extensions would improve robustness to sparsity while preserving OMNIA’s overall design.

Furthermore, we plan to include additional LLM-based baselines beyond our internal prompting variants and to evaluate OMNIA on a broader range of datasets, particularly KGs generated by LLMs, to further validate its adaptability to diverse KG construction contexts. We also intend to refine the filtering component to improve precision while maintaining efficiency, and to explore evaluation metrics better suited to the open-world nature of KGs.

\section*{Acknowledgments}
This work is funded by the French National Agency ANR-21-CE23-0036-01 and he German Research Foundation under the project number 490998901.
\section*{Artifacts}
The code of OMNIA is available on github via the link :\\ https://github.com/fieng94/OMNIA.git
%\bibliography{kg.bib}
%% Please note that all the content must fit within the page limits, including any appendices.
%\appendix
% ...
\section*{AI-Generated Content Acknowledgment}
Some parts of this paper were prepared with the assistance of AI tools. 
These tools were used to support the writing process by improving clarity, coherence, and formatting. 
All scientific ideas, experimental designs, analyses, and interpretations were conceived, validated, and verified by the authors.

\printbibliography
\end{document}